\documentclass[aip,jcp,reprint,floatfix]{revtex4-2}
\usepackage{amsmath,amssymb}
\usepackage{graphicx}
\usepackage{dcolumn}
\usepackage{bm}
\usepackage{multirow}
\usepackage{float}
\usepackage{hhline}
\usepackage{subfigure}
\usepackage{color}
\usepackage{afterpage}
\usepackage{tabularx}
\usepackage{comment}
\usepackage{float}

\usepackage[utf8]{inputenc}
\usepackage[T1]{fontenc}
\usepackage{mathptmx}
\usepackage{physics}
\usepackage{amsmath}

\usepackage[dvipsnames]{xcolor}

\graphicspath{{Figures/}}

\begin{document}

\preprint{AIP/123-QED}

\title{Microscopic theory of exciton-polariton model involving multiple molecules: Macroscopic quantum electrodynamics formulation and essence of direct intermolecular interactions}



\author{Yi-Ting Chuang}
\affiliation{Department of Chemistry, National Taiwan University, Taipei 10617, Taiwan}
\affiliation{Institute of Atomic and Molecular Sciences, Academia Sinica, Taipei 10617, Taiwan}
\author{Liang-Yan Hsu}
\email{lyhsu@gate.sinica.edu.tw}
\affiliation{Department of Chemistry, National Taiwan University, Taipei 10617, Taiwan}
\affiliation{Institute of Atomic and Molecular Sciences, Academia Sinica, Taipei 10617, Taiwan}
\affiliation{Physics Division, National Center for Theoretical Sciences, Taipei 10617, Taiwan}

\begin{abstract}
Cavity quantum electrodynamics (CQED) and its extensions are widely used for the description of exciton-polariton systems. However, the exciton-polariton models based on CQED vary greatly within different contexts. One of the most significant discrepancies among these CQED models is whether one should include direct intermolecular interactions in the CQED Hamiltonian. To answer this question, in this article, we derive an effective dissipative CQED model including free-space dipole-dipole interactions (CQED-DDI) from a microscopic Hamiltonian based on macroscopic quantum electrodynamics. Dissipative CQED-DDI successfully captures the nature of vacuum fluctuations in dielectric media and separates it into the free-space effects and the dielectric-induced effects. The former include spontaneous emissions, dephasings and dipole-dipole interactions in free space; the latter include exciton-polariton interactions and photonic losses due to dielectric media. We apply dissipative CQED-DDI to investigate the exciton-polariton dynamics (the population dynamics of molecules above a plasmonic surface) and compare the results with those based on the methods proposed by several previous studies. We find that direct intermolecular interactions are a crucial element when employing CQED-like models to study exciton-polariton systems involving multiple molecules.

\end{abstract}

\maketitle

\section{Introduction}

Light-matter interaction has long been an important topic in physical chemistry and chemical physics. Recently, the interactions between molecules and confined electromagnetic fields (quantum light) have received considerable attention because within various photonic environments many experiments have shown that quantum light can alter molecular physical and chemical processes, including the collective spontaneous emission \cite{Rohlsberger2010,Goban2015,Hood2016,Zhang2016,Solano2017,Kim2018,Luo2019,Riccardo2022, Tiranov2023}, energy transfer \cite{Andrew2004,Götzinger2006,Choi2009,Lunz2011,Coles2014,Zhong2016,Zhong2017,Xiang2020,Georgiou2021,Wang2021}, and even chemical reactions \cite{Hutchison2012,Thomas2019,Thomas2020,Sau2021,Mony2021,Puro2021,Zeng2023,Ahn2023}. In response to these experimental discoveries, numerous theoretical frameworks have emerged, many of which are based on cavity quantum electrodynamics (CQED) and its extensions. In traditional CQED, molecules are solely coupled to photonic modes, without any direct intermolecular interactions. This convention may be traced back to the Dicke model \cite{Dicke1954} and the Tavis-Cummings model \cite{Tavis1968}, which consider multiple molecules and a single photonic mode. These kinds of models, which do not consider direct intermolecular interactions, have been widely adopted \cite{Agarwal1984,Dung1994,Deng2015,Decordi2018,Ribeiro2018,Evans2018,Angerer2018}. Nonetheless, an alternative form of CQED that incorporates direct molecule-molecule interactions has also been proposed and used \cite{Kurizki1996,Caruso2012,Schachenmayer2015,Zhang2018,Sáez-Blázquez2019,Debnath2020}. The discrepancy between these two formalisms of CQED models gives rise to an open question: \textit{Should one include direct intermolecular interactions in the CQED models when studying exciton-polariton systems involving multiple molecules?}

Indeed, the presence or absence of direct intermolecular interactions in the quantum electrodynamics (QED) Hamiltonian is a well-established result in molecular QED (QED in free space) \cite{Craig1998,Woolley2022}, and it depends upon the chosen theoretical scheme. For instance, in the minimal coupling scheme (in the Coulomb gauge), direct Coulomb interaction terms among different molecules manifest within the Hamiltonian. On the contrary, within the multipolar coupling scheme, also known as the Power-Zienau-Woolley (PZW) framework \cite{Power1959,Woolley1971}, direct molecule-molecule interactions are absent from the Hamiltonian when there is no overlap in the charge distributions of separate molecules. Instead, all interactions are mediated via electromagnetic fields. Given that the Hamiltonians within the multipolar coupling scheme and the minimal coupling scheme are related to each other by a unitary transformation, i.e., the PZW transformation, it follows that any physical quantities computed using these two frameworks should be congruent. Hence, the presence or absence of direct intermolecular interactions in the molecular QED Hamiltonian boils down to a matter of theoretical preference.

Nevertheless, when it comes to CQED, this issue becomes more complicated. Contrary to molecular QED, which encompasses an infinite spectrum of continuous photonic modes, the CQED Hamiltonian is restricted to only a single or a limited number of discrete photonic modes. A recent study indicates that the CQED Hamiltonian with a limited number of photonic modes can easily give misleading results \cite{Pantazopoulos2023}. Given that the direct intermolecular interactions are essentially the combined effect of the full photonic mode spectrum, the CQED Hamiltonian with a few photonic modes generally falls short of fully encapsulating the entirety of these effects. 
In order to prevent us from misinterpreting experimental observations and to accurately describe the light-matter interactions theoretically, clarification of the above issues is urgent and necessary.

To answer the question of whether one should include direct intermolecular interactions in the CQED Hamiltonian, in this work, through a microscopic Hamiltonian based on macroscopic QED (MQED)  \cite{Gruner1996,Dung1998,Scheel1998,Vogel2006b,Buhmann2012b}, we present a theory which allows us to incorporate the free-space dipole-dipole interactions into an effective dissipative CQED model, i.e., CQED with photon loss, and we denote the method as dissipative CQED-DDI. According to the previous studies, MQED not only extends molecular QED into regimes encompassing materials with inhomogeneous, dispersive, and absorbing characteristics \cite{Gruner1996}, but also successfully explains experimental observations in complex dielectric environments \cite{Zhang2014,Wang2019,Boddeti2022}. As a result, we believe that MQED serves as a suitable theoretical framework for accurately describing exciton-polariton systems involving multiple molecules in complex dielectric environments, such as plasmonic surface and Fabry-Pérot cavity.


For the convenience of readers, we briefly outline the main advantages of dissipative CQED-DDI in this work. First, dissipative CQED-DDI clearly delineates that the effects of light-matter interactions in dielectric media can be effectively divided into (i) the free-space effects, including spontaneous emissions, dephasings and dipole-dipole interactions in free space, and (ii) the medium-induced effects, including exciton-polariton(photon) interactions and photonic losses. In fact, dissipative CQED-DDI is an extension of dissipative CQED (without direct intermolecular interactions) that can also be derived from MQED using the few-mode field quantization approach \cite{Sánchez-Barquilla2022}. Second, through a case study, dissipative CQED-DDI successfully captures the effect of direct intermolecular interactions and shows this effect plays an important role in the study of the exciton-polariton systems involving multiple molecules.


Our article is organized as follows. In Sec.~\ref{Sec:MQED}, we recapitulate the MQED theory. In Sec.~\ref{Sec:D-CQED-DDI}, we construct an effective microscopic model based on MQED and derive dissipative CQED-DDI from the effective microscopic model. In Sec.~\ref{Sec:Comparison}, we make a comparison between dissipative CQED-DDI and dissipative CQED which does not include direct intermolecular interactions. In Sec.~\ref{Sec:Numerical Demonstration and Discussion}, through a case study on the exciton-polariton dynamics of
molecules above a plasmonic surface, we demonstrate the advantage of dissipative CQED-DDI against dissipative CQED and reveal the importance of the direct intermolecular interactions in the (dissipative) CQED model. Finally, in Sec.~\ref{Sec:Conclusion}, we summarize the main results of this study.

\section{Method} 

\subsection{Macroscopic quantum electrodynamics}
\label{Sec:MQED}

We consider an ensemble of $N_\mathrm{M}$ two-level molecules coupled to polaritons (dressed photons) in any arbitrary dielectric environment. Based on MQED, the total Hamiltonian of the hybrid light-matter system in the multipolar coupling scheme under the electric-dipole approximation can be expressed as
\begin{align}
    & \hat{H}_\mathrm{T} = \hat{H}_\mathrm{M} + \hat{H}_\mathrm{P}+ \hat{H}_\mathrm{M-P},
    \label{Eq:Hamiltonain_Tot}
\end{align}
with
\begin{align}
    \hat{H}_\mathrm{M} & = \sum^{N_\mathrm{M}}_{\alpha=1} \hbar\omega_\alpha \hat{\sigma}^{(+)}_\alpha \hat{\sigma}^{(-)}_\alpha,
    \label{Eq:Hamiltonian_Mol}\\
    \hat{H}_\mathrm{P} & = \int \mathrm{d}\mathbf{r} \int_{0}^{\infty} \mathrm{d}\omega \, \hbar\omega \,\mathbf{\hat{f}}^\dagger(\bf{r},\omega)\cdot\mathbf{\hat{f}}(\bf{r},\omega),
    \label{Eq:Hamiltonian_Pol}\\
    \hat{H}_\mathrm{M-P} & = - \sum^{N_\mathrm{M}}_{\alpha=1} \hat{\boldsymbol{\mu}}_{\alpha} \cdot \hat{\mathbf{F}}(\mathbf{r}_\alpha).
    \label{Eq:Hamiltonian_Int}
\end{align}
$\hat{H}_\mathrm{M}$ corresponds to the molecular Hamiltonian, where $N_\mathrm{M}$ is the number of molecules, $\omega_\alpha$ and $\hat{\sigma}^{(+)}_\alpha$ ($\hat{\sigma}^{(-)}_\alpha$) are the electronic transition frequency and raising (lowering) operator of $\alpha$ molecule, respectively. $\hat{H}_\mathrm{P}$ corresponds to the polaritonic Hamiltonian, where $\mathbf{\hat{f}}^\dagger(\mathbf{r},\omega)$ and $\mathbf{\hat{f}}(\mathbf{r},\omega)$ are the creation and annihilation operators of the bosonic vector fields \cite{Gruner1996,Dung1998,Scheel1998,Buhmann2012b,Vogel2006b} that obey the commutation relations $\comm*{\hat{f}_k\left( \mathbf{r}, \omega \right)}{\hat{f}^\dagger_{k'}\left( \mathbf{r}', \omega' \right)} = \delta_{kk'} \delta\left( \mathbf{r}  - \mathbf{r}' \right) \delta\left( \omega - \omega' \right)$ and $\comm*{ \hat{f}_k\left( \mathbf{r}, \omega \right)}{\hat{f}_{k'}\left( \mathbf{r}', \omega' \right)} = 0$. $\hat{H}_\mathrm{M-P}$ corresponds to the interaction Hamiltonian, where $\hat{\boldsymbol{\mu}}_{\alpha}$ and $\hat{\mathbf{F}}(\mathbf{r}_\alpha)$ are transition dipole operator of $\alpha$ and the field operator, respectively. The transition dipole operator can be expressed as
\begin{align}
    \hat{\boldsymbol{\mu}}_\alpha = \boldsymbol{\mu}_\alpha \hat{\bar{\mu}}_\alpha,
\end{align}
where $\hat{\bar{\mu}}_\alpha = \hat{\sigma}^{(+)}_{\alpha} + \hat{\sigma}^{(-)}_{\alpha}$, and $\boldsymbol{\mu}_\alpha$ is the value of the transition dipole moment of $\alpha$.
The field operator is expressed as
\begin{align}
    \hat{\mathbf{F}}(\mathbf{r}_\alpha) = \int \dd{\mathbf{r}} \int_0^\infty \dd{\omega}  \overline{\overline{\mathcal{G}}}(\mathbf{r}_\alpha,\mathbf{r},\omega) \cdot \hat{\mathbf{f}}(\mathbf{r},\omega) + \mathrm{H.c.},
\end{align}
with
\begin{gather}
    \overline{\overline{\mathcal{G}}}(\mathbf{r}_\alpha,\mathbf{r},\omega) = i\sqrt{\frac{\hbar}{\pi\varepsilon_0}} \frac{\omega^2}{c^2} \sqrt{\mathrm{Im} \left[ \varepsilon_\mathrm{r}(\mathbf{r},\omega) \right]} \, \overline{\overline{\mathbf{G}}}(\mathbf{r}_\alpha,\mathbf{r},\omega).
\end{gather}
$\varepsilon_0$ and $\varepsilon_\mathrm{r}(\mathbf{r},\omega)$ are the permittivity of free space and position-dependent relative permittivity, respectively; $c$ is the speed of light in vacuum. $\overline{\overline{\mathcal{G}}}(\mathbf{r}_\alpha,\mathbf{r},\omega)$ is an auxiliary tensor related to the dyadic Green's function $\overline{\overline{\mathbf{G}}}(\mathbf{r},\mathbf{r'},\omega)$ which satisfies macroscopic Maxwell’s equations, i.e.,
\begin{equation}
    \left[ \frac{\omega^2}{c^2}\varepsilon_\mathrm{r}(\mathbf{r},\omega) - \nabla \times \nabla \times \right] \overline{\overline{\mathbf{G}}}(\mathbf{r},\mathbf{r'},\omega) = -\mathbf{\overline{\overline{I}}}_3 \delta(\mathbf{r}-\mathbf{r'}),
\label{Eq:GreensFunction}
\end{equation}
where $\mathbf{\overline{\overline{I}}}_3$ is the $3\times3$ identity matrix, and $\delta(\mathbf{r}-\mathbf{r'})$ is the three-dimensional delta function. The dyadic Green's function can be further decomposed as \cite{Buhmann2012b}
\begin{align}
    \overline{\overline{\mathbf{G}}}(\mathbf{r},\mathbf{r}',\omega) = \overline{\overline{\mathbf{G}}}_0(\mathbf{r},\mathbf{r}',\omega) + \overline{\overline{\mathbf{G}}}_\mathrm{Sc}(\mathbf{r},\mathbf{r}',\omega),
\label{Eq:G}
\end{align}
where  $\overline{\overline{\mathbf{G}}}_0(\mathbf{r},\mathbf{r}',\omega)$ is the free-space (or bulk) dyadic Green's function of an infinitely extended homogeneous medium, which is free space in our study, and $\overline{\overline{\mathbf{G}}}_\mathrm{Sc}(\mathbf{r},\mathbf{r}',\omega)$ is the scattering dyadic Green's function that originates from the presence of the dielectric bodies.

For the hybrid light-matter system described by the MQED Hamiltonian in Eq.~(\ref{Eq:Hamiltonain_Tot}), we can define its generalized spectral density $\overline{\overline{\mathbf{J}}}(\omega)$ as \cite{Sánchez-Barquilla2022,Chuang2022}
\begin{align}
    J_{\alpha \beta}(\omega) & = \frac{\omega^2}{\pi \hbar \epsilon_0 c^2} \boldsymbol{\mu}_\alpha \cdot \mathrm{Im} \overline{\overline{\mathbf{G}}}(\mathbf{r}_\alpha, \mathbf{r}_\beta,\omega) \cdot \boldsymbol{\mu}_\beta.
\end{align}
The generalized spectral density encapsulates comprehensive details regarding the interactions between the molecules and the polaritons. Consequently, the molecules will exhibit consistent dynamics when they are placed in varied photonic environments that are governed by an identical $\overline{\overline{\mathbf{J}}}(\omega)$. Furthermore, one can create diverse models to simulate the dynamics of the molecules with higher efficiency by setting the generalized spectral density of the models to align with $\overline{\overline{\mathbf{J}}}(\omega)$. According to Eq.~(\ref{Eq:G}), the generalized spectral density $\overline{\overline{\mathbf{J}}}(\omega)$ can be further decomposed as the sum of the free-space generalized spectral density $\overline{\overline{\mathbf{J}}}_0(\omega)$ and the scattering generalized spectral density $\overline{\overline{\mathbf{J}}}_\mathrm{Sc}(\omega)$, where
\begin{subequations}
\begin{align}
    J_{0, \alpha \beta}(\omega) & = \frac{\omega^2}{\pi \hbar \epsilon_0 c^2} \boldsymbol{\mu}_\alpha \cdot \mathrm{Im} \overline{\overline{\mathbf{G}}}_0(\mathbf{r}_\alpha, \mathbf{r}_\beta,\omega) \cdot \boldsymbol{\mu}_\beta, \\
    J_{\mathrm{Sc}, \alpha \beta}(\omega) & = \frac{\omega^2}{\pi \hbar \epsilon_0 c^2} \boldsymbol{\mu}_\alpha \cdot \mathrm{Im} \overline{\overline{\mathbf{G}}}_\mathrm{Sc}(\mathbf{r}_\alpha, \mathbf{r}_\beta,\omega) \cdot \boldsymbol{\mu}_\beta.
\end{align}    
\end{subequations}

Here, we would like to clarify that "macroscopic" QED refers to the quantization of macroscopic Maxwell's equations (which corresponds to polariton or photon dressed by dielectric medium); therefore, despite polaritons (dressed photons) originating from macroscopic dielectric functions, the MQED theory can be still regarded as a microscopic description of exciton-polariton systems.

\subsection{Dissipative cavity quantum electrodynamics including free-space dipole-dipole interactions}
\label{Sec:D-CQED-DDI}

In this section, we aim to construct a dissipative CQED model including free-space dipole-dipole interactions (CQED-DDI) that mirrors the quantum dynamics of the molecules as described in the MQED theory. To derive dissipative CQED-DDI, we begin with an effective microscopic model $\hat{\mathcal{H}}_\mathrm{T}$ which possesses an identical generalized spectral density as the MQED Hamiltonian. We separate the effective microscopic model into the system part, the bath part, and the system-bath interaction. The bath part is then traced out, yielding dissipative CQED-DDI.

\subsubsection{Effective microscopic model}

The effective microscopic model $\hat{\mathcal{H}}_\mathrm{T}$ is defined as
\begin{align}
    \hat{\mathcal{H}}_\mathrm{T} = \hat{H}_\mathrm{M} + \hat{\mathcal{H}}^0_\mathrm{Ph} + \hat{\mathcal{H}}^0_\mathrm{M-Ph} + \hat{\mathcal{H}}_\mathrm{Ph} + \hat{\mathcal{H}}_\mathrm{M-Ph}.
\label{Eq:Microscopic_Model}
\end{align}
Within this model, the molecules $\hat{H}_\mathrm{M}$ are coupled to $\hat{\mathcal{H}}^0_\mathrm{Ph}$ and $\hat{\mathcal{H}}_\mathrm{Ph}$. Since $\hat{\mathcal{H}}^0_\mathrm{Ph}$ and $\hat{\mathcal{H}}_\mathrm{Ph}$ will account for the free-space effects and the dielectric-induced effects, respectively, we denote them as the free-space photonic modes and the scattering photonic modes. The molecular Hamiltonian $\hat{H}_\mathrm{M}$ is identical to that in the MQED Hamiltonian in Eq.~(\ref{Eq:Hamiltonian_Mol}). The free-space photonic mode Hamiltonian $\hat{\mathcal{H}}^0_\mathrm{Ph}$ comprises $N_\mathrm{M}$ independent continuous photonic modes, i.e.,
\begin{align}
    \hat{\mathcal{H}}^0_\mathrm{ph} &= \sum_{l=1}^{N_\mathrm{M}} \int_{-\infty}^{\infty} \dd{\omega} \hbar \omega \hat{c}^\dagger_l(\omega) \hat{c}_l(\omega),
\end{align}
where $\hat{c}_l (\omega)$ ($\hat{c}^\dagger_{l} (\omega')$) obeys the commutation relations $\comm*{\hat{c}_l (\omega)}{\hat{c}^\dagger_{l'} (\omega')} = \delta_{ll'}\delta\left(\omega -\omega'\right)$ and $\comm*{\hat{c}_l (\omega)}{\hat{c}_{l'} (\omega')} = 0$. The interactions between the molecules and free-space photonic modes are defined as
\begin{align}
    \hat{\mathcal{H}}^0_\mathrm{int} &= \sum_{\alpha, l} \int_{-\infty}^{\infty} \dd{\omega} \hbar g_{\alpha l}(\omega) \hat{\bar{\mu}}_\alpha \left[ \hat{c}^\dagger_l(\omega) + \hat{c}_l(\omega) \right],
\end{align}
where the interaction strength between the $\alpha$-th molecule and the $l$-th free-space photonic mode $g_{\alpha l}(\omega)$ is set to be \cite{Feist2021}
\begin{align}
    g_{\alpha l}(\omega) = \theta(\omega) \sqrt{J_{0, \alpha \alpha}(\omega)} W_{\alpha l}(\omega).
\label{Eq:g_MPh0}
\end{align}
$\theta(\omega)$ is the Heaviside step function, and $\overline{\overline{\mathbf{W}}} (\omega)$ is a $N_\mathrm{M} \times N_\mathrm{M}$
square matrix that satisfies
\begin{align}
    \overline{\overline{\mathbf{W}}} (\omega) \overline{\overline{\mathbf{W}}}\vphantom{a}^\mathrm{T} (\omega) = \overline{\overline{\mathbf{S}}} (\omega), 
\label{Eq:M}
\end{align}
with
\begin{align}
    S_{\alpha \beta}(\omega) = \frac{J_{0, \alpha \beta}(\omega)}{\sqrt{J_{0, \alpha \alpha}(\omega) J_{0, \beta \beta}(\omega)}}.
\label{Eq:S}
\end{align}

The scattering photonic mode Hamiltonian $\hat{\mathcal{H}}_\mathrm{ph}$ comprises a collection of discrete primary scattering photonic modes $\hat{\mathcal{H}}_\mathrm{ph1}$, which in turns couple to a collection of independent continuous Markovian secondary scattering photonic modes $\hat{\mathcal{H}}_\mathrm{ph2}$,
\begin{align}
    & \hat{\mathcal{H}}_\mathrm{Ph} = \hat{\mathcal{H}}_\mathrm{Ph1} + \hat{\mathcal{H}}_\mathrm{Ph2} + \hat{\mathcal{H}}_\mathrm{Ph1-Ph2},
\end{align}
with
\begin{align}
    \hat{\mathcal{H}}_\mathrm{Ph1} & = \sum^{N_\mathrm{ph}}_{j=1} \hbar \omega_{\mathrm{ph}, j} \hat{a}^\dagger_j \hat{a}_j, \\
    \hat{\mathcal{H}}_\mathrm{Ph2} & = \sum^{N_\mathrm{ph}}_{j=1} \int_{-\infty}^\infty \dd{\omega} \hbar \omega \hat{b}_j^\dagger(\omega) \hat{b}_j(\omega), \\
    \hat{\mathcal{H}}_\mathrm{Ph1-Ph2} & = \sum^{N_\mathrm{ph}}_{j=1} \int_{-\infty}^\infty \dd{\omega} \hbar\sqrt{\frac{\kappa_{\mathrm{ph}, j}}{2\pi}} \left[ \hat{b}_j^\dagger(\omega) \hat{a}_j + \hat{b}_j(\omega) \hat{a}_j^\dagger \right],
\label{EQ:Hamiltonian_Ph1-Ph2}
\end{align}
where $N_\mathrm{ph}$ is the number of the discrete primary scattering photonic modes and has not been determined yet; $\omega_{\mathrm{ph}, j}$ is the frequency of the $j$-th discrete primary 
scattering photonic mode; $\hat{a}^\dagger_j$ ($\hat{a}_j$) is the creation (annihilation) operator of the $j$-th discrete primary scattering photonic mode that obeys the commutation relations $\comm*{\hat{a}_j}{\hat{a}^\dagger_{j'}} = \delta_{jj'}$ and $\comm*{\hat{a}_j}{\hat{a}_{j'}}  = 0$. Note that the primary scattering photonic modes can be generalized to the interacting modes \cite{Medina2021,Sánchez-Barquilla2022} i.e., $\hat{\mathcal{H}}_\mathrm{Ph1} = \sum_{i,j} \hbar \omega_{\mathrm{ph}, ij} \hat{a}^\dagger_i \hat{a}_j$, if required. $\hat{b}_j(\omega)$ ($\hat{b}_j^\dagger(\omega)$) is the creation (annihilation) operator of the $j$-th continuous secondary scattering photonic mode that obeys the commutation relations $\comm*{\hat{b}_j (\omega)}{\hat{b}^\dagger_{j'} (\omega')} = \delta_{jj'}\delta\left(\omega -\omega'\right)$ and $\comm*{\hat{b}_j (\omega)}{\hat{b}_{j'} (\omega')} = 0$; $\hbar\sqrt{\kappa_{\mathrm{ph}, j}/2\pi}$ is the interaction strength between the $j$-th discrete primary scattering photonic mode $\hat{a}_j$ and its own Markovian reservoir (continuous secondary scattering photonic mode) $\hat{b}_j(\omega)$. For the interactions between the molecules and the scattering photonic modes, the molecules are only coupled to the discrete primary scattering photonic modes, i.e.,
\begin{align}
    \hat{\mathcal{H}}_\mathrm{M-Ph} &= \sum_{\alpha, j} \hbar \Omega_{\alpha j} \hat{\bar{\mu}}_\alpha \left( \hat{a}^\dagger_j + \hat{a}_j \right),
\end{align}
where $\hbar \Omega_{\alpha j}$ is the interaction strength between the $\alpha$-th molecule and $j$-th primary scattering photonic mode.

The generalized spectral density $\overline{\overline{\mathfrak{J}}}(\omega)$ of the effective microscopic model can be expressed as
\begin{align}
    \overline{\overline{\mathfrak{J}}}(\omega) = \overline{\overline{\mathfrak{J}}}_0(\omega) + \overline{\overline{\mathfrak{J}}}_\mathrm{Sc}(\omega),
\end{align}
where $\overline{\overline{\mathfrak{J}}}_0(\omega)$ accounts for the interactions between the molecules and the free-space photonic modes, and $\overline{\overline{\mathfrak{J}}}_\mathrm{Sc}(\omega)$ accounts for the interactions between the molecules and the scattering photonic modes. $\overline{\overline{\mathfrak{J}}}_0(\omega)$ is given by \cite{Sánchez-Barquilla2022}
\begin{align}
    \overline{\overline{\mathfrak{J}}}_0(\omega) & = \overline{\overline{\mathbf{g}}}(\omega) \overline{\overline{\mathbf{g}}}\vphantom{a}^\mathrm{T}(\omega),
\end{align}
where $g_{\alpha j}$ is defined in Eq.~(\ref{Eq:g_MPh0}); $\overline{\overline{\mathfrak{J}}}_\mathrm{Sc}(\omega)$ is given by \cite{Sánchez-Barquilla2022}
\begin{align}
    \overline{\overline{\mathfrak{J}}}_\mathrm{Sc}(\omega) = \frac{1}{\pi} \mathrm{Im} \left[  \overline{\overline{\boldsymbol{\Omega}}} \cdot   (\overline{\overline{\mathcal{H}}}_\mathrm{eff} - \omega  \overline{\overline{\mathbf{I}}}_{N_\mathrm{ph}})^{-1} \cdot  \overline{\overline{\boldsymbol{\Omega}}}\vphantom{a}^\mathrm{T} \right],
\end{align}
where $\mathcal{H}_{\mathrm{eff},jk} = (\omega_{\mathrm{ph}, j} - \frac{i}{2} \kappa_{\mathrm{ph}, j}) \delta_{jk}$, and $\overline{\overline{\mathbf{I}}}_{N_\mathrm{ph}}$ is a $N_\mathrm{ph} \times N_\mathrm{ph}$ identity matrix. Note that $\overline{\overline{\mathfrak{J}}}_0(\omega)$ is indeed identical to $\overline{\overline{\mathbf{J}}}_\mathrm{0}(\omega)$; therefore, the only plausible scenario where the effective microscopic model possesses an identical generalized spectral density as the MQED Hamiltonian is by ensuring the identity of $\overline{\overline{\mathfrak{J}}}_\mathrm{Sc}(\omega)$ with $\overline{\overline{\mathbf{J}}}_\mathrm{Sc}(\omega)$. This implies that the parameters $N_\mathrm{ph}$, $\omega_\mathrm{ph,j}$, $\Omega_{\alpha j}$ and $\kappa_{\mathrm{ph}, j}$ are obtainable by fitting the scattering generalized spectral density of the MQED Hamiltonian $\overline{\overline{\mathbf{J}}}_\mathrm{Sc}(\omega)$ with its counterpart in the effective microscopic model $\overline{\overline{\mathfrak{J}}}_\mathrm{Sc}(\omega)$.

\subsubsection{System-bath Hamiltonian and dissipative CQED-DDI}

Finally, to derive the equation of motion of the molecules, we introduce the concept of the system-bath Hamiltonian \textcolor{black}{and separate the total Hamiltonian $\hat{\mathcal{H}}_\mathrm{T}$ into the system Hamiltonian $\hat{\mathcal{H}}_\mathrm{S}$, the bath Hamiltonian $\hat{\mathcal{H}}_\mathrm{B}$, the system-bath interaction Hamiltonian $\hat{\mathcal{H}}_\mathrm{S-B}$, i.e.,} 
\begin{align}
    \hat{\mathcal{H}}_\mathrm{T} = \hat{\mathcal{H}}_\mathrm{S} + \hat{\mathcal{H}}_\mathrm{B} + \hat{\mathcal{H}}_\mathrm{S-B}.
\label{Eq:Hamiltonian_SB}
\end{align}
\textcolor{black}{The key ideas of how to separate the system and the bath are summarized as follows. First, since the light-matter interactions in free space are typically weak, we neglect their memory effects and treat the free-space photonic modes $\hat{\mathcal{H}}^0_\mathrm{Ph}$ as Markovian baths. Second, the light-matter interactions mediated by the dielectric environments can be either weak or strong. Therefore, to adequately deal with the dielectric effects, we separate the scattering photonic modes into the discrete primary scattering photonic modes $\hat{\mathcal{H}}_\mathrm{Ph1}$, continuous secondary scattering photonic modes $\hat{\mathcal{H}}_\mathrm{Ph2}$, and their couplings $\hat{\mathcal{H}}_\mathrm{Ph1-Ph2}$. Furthermore, we consider the molecules and the primary scattering photonic modes together as the system; hence, the primary scattering photonic modes can account for the non-Markovian effects induced by the dielectric environments. The rest of the scattering photonic modes, i.e., the continuous secondary scattering photonic modes, are designed to be spectrally
flat, and it is reasonable to treat these modes as Markovian baths. According to the above ideas, we design $\hat{\mathcal{H}}_\mathrm{S}$, $\hat{\mathcal{H}}_\mathrm{B}$, and $\hat{\mathcal{H}}_\mathrm{S-B}$ as}
\begin{align}
    \hat{\mathcal{H}}_\mathrm{S} & = \hat{H}_\mathrm{M} + \hat{\mathcal{H}}_\mathrm{Ph1} + \hat{\mathcal{H}}_\mathrm{M-Ph}, \\
    \hat{\mathcal{H}}_\mathrm{B} & = \hat{\mathcal{H}}^0_\mathrm{Ph} + \hat{\mathcal{H}}_\mathrm{Ph2}, \\
    \hat{\mathcal{H}}_\mathrm{S-B} & = \hat{\mathcal{H}}^0_\mathrm{M-Ph} + \hat{\mathcal{H}}_\mathrm{Ph1-Ph2}.
\end{align}

Since the bath degrees of freedom are all Markovian, we can simply trace them out. \textcolor{black}{In addition, we apply the initial condition that the bath modes (the photonic modes) are initially in the vacuum state, i.e., the bath is assumed to be at zero temperature.} As a result, we obtain the master equation of the system density matrix $\hat{\rho}_\mathrm{S}(t)$ as (see Appendix~\ref{Sec:AppendixA} for the details)
\begin{align}
\nonumber
    & \pdv{t} \hat{\rho}_\mathrm{S}(t) =  -\frac{i}{\hbar} \left[ \hat{\mathcal{H}}_\mathrm{S} + \hat{\mathcal{H}}^0_\mathrm{ES}, \hat{\rho}_\mathrm{S}(t) \right] \\
\nonumber  
    & \quad + \sum_{\alpha, \beta} \Gamma^0_{\alpha \beta} \left[ \hat{\sigma}^{(-)}_\beta \hat{\rho}_\mathrm{S}(t) \hat{\sigma}^{(+)}_\alpha - \frac{1}{2} \left\{ \hat{\sigma}^{(+)}_\alpha \hat{\sigma}^{(-)}_\beta, \hat{\rho}_\mathrm{S}(t) \right\} \right] \\
\nonumber  
    & \quad + \sum_{\alpha, \beta} \gamma^0_{\alpha \beta} \left[ \hat{\sigma}^{(+)}_\beta \hat{\rho}_\mathrm{S}(t) \hat{\sigma}^{(-)}_\alpha - \frac{1}{2} \left\{ \hat{\sigma}^{(-)}_\alpha \hat{\sigma}^{(+)}_\beta, \hat{\rho}_\mathrm{S}(t) \right\} \right] \\  
    & \quad  + \sum_j \kappa_{\mathrm{ph}, j} \left[ \hat{a}_j \hat{\rho}_\mathrm{S}(t) \hat{a}^\dagger_j - \frac{1}{2} \left\{ \hat{a}^\dagger_j \hat{a}_j, \hat{\rho}_\mathrm{S}(t) \right\} \right],
    \label{Eq:EOM_rho_CQED}
\end{align}
where $\left\{ \hat{O}_1, \hat{O}_2 \right\} = \hat{O}_1 \hat{O}_2 + \hat{O}_2 \hat{O}_1$ is the anticommutator. Recall that $\kappa_{\mathrm{ph}, j}$ has been defined below Eq.~(\ref{EQ:Hamiltonian_Ph1-Ph2}). \textcolor{black}{$\Gamma^0_{\alpha \beta}$ is associated with the free-space molecular dissipation (including spontaneous emission and dephasing) rate, and $\gamma^0_{\alpha \beta}$ is associated with the free-space dephasing rate due to counter-rotating interactions; they are expressed as}
\begin{align}
    \Gamma^0_{\alpha \beta} & = \pi \left[ J_{0, \alpha \beta}(\omega_\beta) + J_{0, \beta \alpha}(\omega_\alpha) \right] \\
\nonumber
    & \quad - i \left[ \delta^{0-}_{\alpha\beta}(\omega_\beta) - \delta^{0-}_{\beta\alpha}(\omega_\alpha) \right], \\
    \gamma^0_{\alpha \beta} & = - i \left[ \delta^{0+}_{\alpha\beta}(\omega_\beta) - \delta^{0+}_{\beta\alpha}(\omega_\alpha) \right], 
\end{align}
with
\begin{align}
    \delta^{0\pm}_{\alpha\beta}(\omega) &= \mathcal{P} \int_0^\infty \dd{\omega'} \frac{J^0_{\alpha \beta}(\omega')}{\omega' \pm \omega}.
\end{align}
In addition, $\hat{\mathcal{H}}^0_\mathrm{ES}$ in Eq.~(\ref{Eq:EOM_rho_CQED}) is the effective free-space energy shift Hamiltonian (also known as the Lamb-type Hamiltonian) induced by the system(molecules)-bath(free-space photonic modes) interactions. $\hat{\mathcal{H}}^0_\mathrm{ES}$ includes both the free-space Lamb shift Hamiltonian $\hat{\mathcal{H}}^0_\mathrm{LS}$ (diagonal energy shifts) and the free-space dipole-dipole interaction Hamiltonian $\hat{\mathcal{H}}^0_\mathrm{DDI}$ (off-diagonal couplings), i.e.,
\begin{align}
    \hat{\mathcal{H}}^0_\mathrm{ES} & = \hat{\mathcal{H}}^0_\mathrm{LS} + \hat{\mathcal{H}}^0_\mathrm{DDI}, \\
    \hat{\mathcal{H}}^0_\mathrm{LS} & = \sum^{N_\mathrm{M}}_{\alpha=1} \Delta^0_\alpha \hat{\sigma}^{(+)}_\alpha \hat{\sigma}^{(-)}_\alpha, \\
    \hat{\mathcal{H}}^0_\mathrm{DDI} & = \sum_{\substack{\alpha,\beta\\
    \alpha\neq\beta}} V^0_{\alpha \beta} \hat{\sigma}^{(+)}_\alpha \hat{\sigma}^{(-)}_\beta,
\end{align}
where $\Delta^0_\alpha$ denotes the free-space Lamb shift, i.e.,
\begin{align}
    \Delta^0_\alpha & = - \hbar \left[ \delta^{0-}_{\alpha\alpha}(\omega_\alpha) - \delta^{0+}_{\alpha\alpha}(\omega_\alpha) \right], 
    \end{align}
and $V^0_{\alpha \beta}$ denotes the free-space dipole-dipole interaction, i.e.,    
\begin{align}
\nonumber
    V^0_{\alpha \beta} & = - \frac{\hbar}{2} \left[ \delta^{0-}_{\alpha\beta}(\omega_\beta) + \delta^{0+}_{\alpha\beta}(\omega_\beta) + i \pi J_{0, \alpha \beta}(\omega_\beta) \right. \\    
    & \left. \quad \quad \quad + \delta^{0-}_{\beta\alpha}(\omega_\alpha) + \delta^{0+}_{\beta\alpha}(\omega_\alpha) - i \pi J_{0, \beta \alpha}(\omega_\alpha) \right] \\
\nonumber    
    & = - \frac{1}{2} \left[ \frac{\omega_\beta^2}{\varepsilon_0 c^2} \boldsymbol{\mu}_\alpha \cdot  \overline{\overline{\mathbf{G}}}_0(\mathbf{r}_\alpha, \mathbf{r}_\beta,\omega_\beta) \cdot \boldsymbol{\mu}_\beta \right. \\
    & \quad \quad \quad \left. + \frac{\omega_\alpha^2}{\varepsilon_0 c^2} \boldsymbol{\mu}_\beta \cdot  \overline{\overline{\mathbf{G}}}\vphantom{a}^*_0(\mathbf{r}_\beta, \mathbf{r}_\alpha,\omega_\alpha) \cdot \boldsymbol{\mu}_\alpha \right].
\end{align}

The free-space Lamb shifts in the master equation can be neglected because their contribution to the energy shift is typically small (after renormalization) \cite{Power1959}, i.e.,
\begin{align}
    \Delta^0_\alpha \approx 0.
\label{Eq:Approx_Lamb}
\end{align}
In addition, since the free-space generalized spectral density is slowly varying (spectrally flat), we can make the substitution $\omega_{\alpha(\beta)} \rightarrow \bar{\omega}_{\alpha \beta} = (\omega_{\alpha} + \omega_{\beta}) / 2$ in $\Gamma^0_{\alpha \beta}$, $\gamma^0_{\alpha \beta}$, and $V^0_{\alpha \beta}$ (which is valid even for relatively large differences in the molecular transition frequencies \cite{Dung2002_2}), resulting in
\begin{align}  
    \tilde{\Gamma}^0_{\alpha \beta} & \approx \frac{2 \bar{\omega}_{\alpha\beta}^2}{\hbar \epsilon_0 c^2} \boldsymbol{\mu}_\alpha \cdot \mathrm{Im} \overline{\overline{\mathbf{G}}}_0(\mathbf{r}_\alpha, \mathbf{r}_\beta,\bar{\omega}_{\alpha\beta}) \cdot \boldsymbol{\mu}_\beta, \\
    \tilde{\gamma}^0_{\alpha \beta} & \approx 0, \\
    \tilde{V}^0_{\alpha \beta} & \approx - \frac{\bar{\omega}_{\alpha\beta}^2}{\varepsilon_0 c^2} \boldsymbol{\mu}_\alpha \cdot  \mathrm{Re}\overline{\overline{\mathbf{G}}}_0(\mathbf{r}_\alpha, \mathbf{r}_\beta,\bar{\omega}_{\alpha\beta}) \cdot \boldsymbol{\mu}_\beta,
\label{Eq:Approx_DDI}
\end{align}
where we have used the identity $\boldsymbol{\mu}_\alpha \cdot  \overline{\overline{\mathbf{G}}}(\mathbf{r}_\alpha, \mathbf{r}_\beta,\omega) \cdot \boldsymbol{\mu}_\beta = \boldsymbol{\mu}_\beta \cdot  \overline{\overline{\mathbf{G}}}(\mathbf{r}_\beta, \mathbf{r}_\alpha,\omega) \cdot \boldsymbol{\mu}_\alpha$. Inserting $\overline{\overline{\mathbf{G}}}_0(\mathbf{r}_\alpha, \mathbf{r}_\beta,\omega)$ \cite{Buhmann2012b} into $\tilde{\Gamma}^0_{\alpha \beta}$ and $\tilde{V}^0_{\alpha \beta}$, one can show that $\tilde{\Gamma}^0_{\alpha \beta}$ is exactly the dissipation (or damping) rate in free space, encompassing the spontaneous emission rate when $\alpha=\beta$ and dephasing rate when $\alpha\neq\beta$, and $\tilde{V}^0_{\alpha \beta}$ ($\alpha\neq\beta$) corresponds exactly to the resonant dipole-dipole interaction in free space \cite{Stephen1964,Craig1998}.

\begin{figure}[!t]
    \centering
    \includegraphics[width=0.48\textwidth]{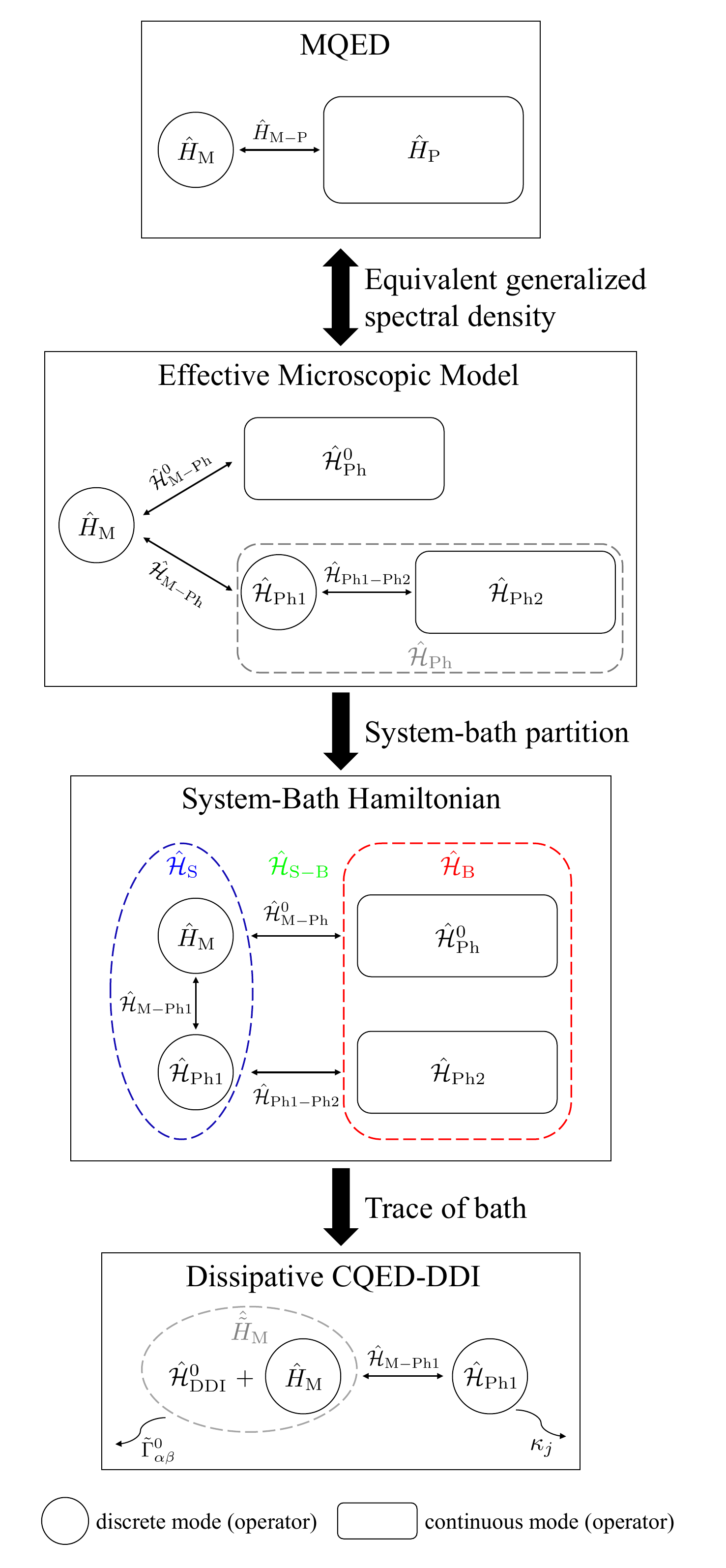}
    \caption{Derivation scheme of dissipative CQED-DDI. The MQED Hamiltonian, the effective microscopic model, the system-bath Hamiltonian and dissipative CQED-DDI are defined in Eq.~(\ref{Eq:Hamiltonain_Tot}), Eq.~(\ref{Eq:Microscopic_Model}), Eq.~(\ref{Eq:Hamiltonian_SB}) and Eq.~(\ref{Eq:EOM_rho2_CQED}), respectively.}
    \label{Fig1}
\end{figure}

After making the approximations in Eqs.~(\ref{Eq:Approx_Lamb}) to (\ref{Eq:Approx_DDI}), the master equation in Eq.~(\ref{Eq:EOM_rho_CQED}) can be simplified as
\begin{align}
\nonumber
    & \pdv{t} \hat{\rho}_\mathrm{S}(t) =  -\frac{i}{\hbar} \left[ \hat{\tilde{H}}_\mathrm{M} + \hat{\mathcal{H}}_\mathrm{Ph1} + \hat{\mathcal{H}}_\mathrm{M-Ph}, \hat{\rho}_\mathrm{S}(t) \right] \\
\nonumber  
    & \quad + \sum_{\alpha, \beta} \tilde{\Gamma}^0_{\alpha \beta} \left[ \hat{\sigma}^{(-)}_\beta \hat{\rho}_\mathrm{S}(t) \hat{\sigma}^{(+)}_\alpha - \frac{1}{2} \left\{ \hat{\sigma}^{(+)}_\alpha \hat{\sigma}^{(-)}_\beta, \hat{\rho}_\mathrm{S}(t) \right\} \right] \\
    & \quad  + \sum_j \kappa_{\mathrm{ph}, j} \left[ \hat{a}_j \hat{\rho}_\mathrm{S}(t) \hat{a}^\dagger_j - \frac{1}{2} \left\{ \hat{a}^\dagger_j \hat{a}_j, \hat{\rho}_\mathrm{S}(t) \right\} \right],
    \label{Eq:EOM_rho2_CQED}
\end{align}
where
\begin{align}
    \hat{\tilde{H}}_\mathrm{M} & = \hat{H}_\mathrm{M} + \hat{\mathcal{H}}^0_\mathrm{DDI}, \\
    \hat{\mathcal{H}}^0_\mathrm{DDI} & = \sum_{\substack{\alpha,\beta\\
    \alpha\neq\beta}} \tilde{V}^0_{\alpha \beta} \hat{\sigma}^{(+)}_\alpha \hat{\sigma}^{(-)}_\beta.
\end{align}
Note that Eq.~(\ref{Eq:EOM_rho2_CQED}) is the main result of this work (the equation of motion for dissipative CQED-DDI). According to Eq.~(\ref{Eq:EOM_rho2_CQED}), the quantum dynamics of an ensemble of molecules in any arbitrary dielectric environment can be equivalently described by a dissipative CQED model including free-space dipole-dipole interactions. The free-space dipole-dipole interactions $\tilde{V}^0_{\alpha \beta}$ and molecular dissipation (including spontaneous emission and dephasing) rates $\tilde{\Gamma}^0_{\alpha \beta}$ can be evaluated in terms of the free-space dyadic Green's functions $\overline{\overline{\mathbf{G}}}_0(\mathbf{r}, \mathbf{r}', \omega)$; the frequencies of the photonic modes $\omega_{\mathrm{ph}, j}$, the photonic loss rates $\kappa_{\mathrm{ph}, j}$, and the molecule-photon (exciton-polariton) coupling strengths $\Omega_{\alpha j}$ can be obtained by fitting the scattering generalized spectral density $\overline{\overline{\mathbf{J}}}_\mathrm{Sc}(\omega)$ to an effective scattering generalized spectral density $\overline{\overline{\mathfrak{J}}}_\mathrm{Sc}(\omega)$. In the end, We summarize the derivation scheme of dissipative CQED-DDI in FIG.~\ref{Fig1}.

\section{Comparison between dissipative CQED-DDI and dissipative CQED}
\label{Sec:Comparison}

\begin{figure*}[!t]
    \centering
    \includegraphics[width=1\textwidth]{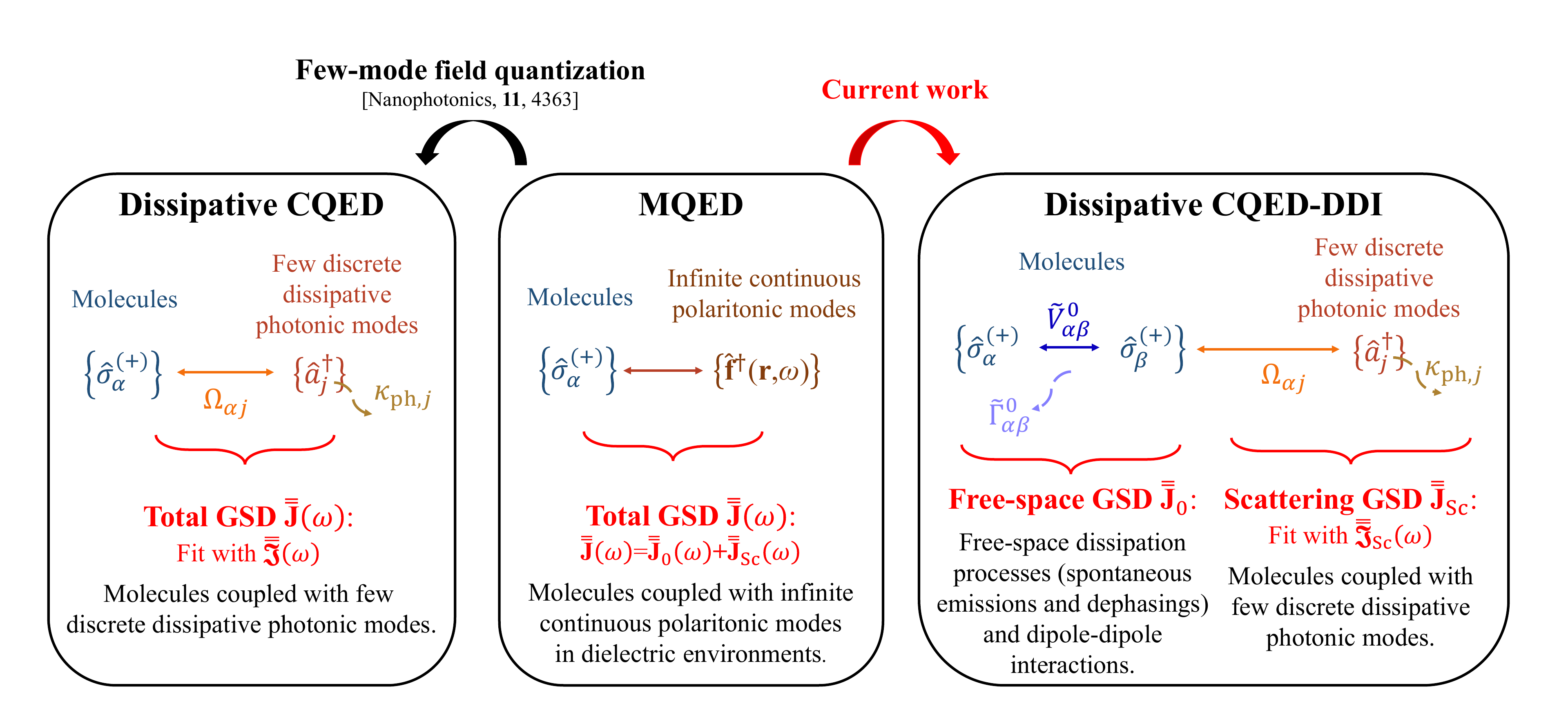}
    \caption{Comparison between MQED, dissipative CQED-DDI in our current work, and dissipative CQED derived from the few-mode approach \cite{Sánchez-Barquilla2022}.}
    \label{Fig2}
\end{figure*}

In the previous section, we have shown how to derive dissipative CQED-DDI from an effective microscopic model based on MQED. In this section, we will compare dissipative CQED-DDI with dissipative CQED, which does not include any direct intermolecular interactions and can also be derived from the MQED theory based on the few-mode field quantization approach \cite{Sánchez-Barquilla2022}. \textcolor{black}{The key difference between dissipative CQED-DDI and dissipative CQED is how to separate the photonic modes. Mathematically, in dissipative CQED-DDI, we identify the free-space photonic modes from the free-space generalized spectral density $\overline{\overline{\mathbf{J}}}_0(\omega)$ and the primary (discrete) and secondary (continuous) scattering photonic modes from the scattering generalized spectral density $\overline{\overline{\mathbf{J}}}_{\mathrm{Sc}}(\omega)$; in dissipative CQED, the previous study only identified the primary and secondary photonic modes from the total generalized spectral density $\overline{\overline{\mathbf{J}}}(\omega)=\overline{\overline{\mathbf{J}}}_{\mathrm{0}}(\omega)+\overline{\overline{\mathbf{J}}}_{\mathrm{Sc}}(\omega)$ without separating them into the free-space and scattering contributions.}

The derivation of dissipative CQED from the MQED theory is similar to that of dissipative CQED-DDI. We begin with an effective microscopic model $\hat{\mathcal{H}}'_\mathrm{T}$, where
\begin{align}
    \hat{\mathcal{H}}'_\mathrm{T} &= \hat{H}_\mathrm{M} + \hat{\mathcal{H}}'_\mathrm{Ph} + \hat{\mathcal{H}}'_\mathrm{M-Ph},
\end{align}
with
\begin{align}
    \hat{\mathcal{H}}'_\mathrm{Ph} &= \hat{\mathcal{H}}'_\mathrm{Ph1} + \hat{\mathcal{H}}'_\mathrm{Ph2} + \hat{\mathcal{H}}'_\mathrm{Ph1-Ph2}, \\
    \hat{\mathcal{H}}'_\mathrm{Ph1} & = \sum^{N'_\mathrm{ph}}_{j=1} \hbar \omega'_{\mathrm{ph}, j} \hat{a}'^\dagger_j \hat{a}'_j, \\
    \hat{\mathcal{H}}'_\mathrm{Ph2} & = \sum^{N'_\mathrm{ph}}_{j=1} \int_{-\infty}^\infty \dd{\omega} \hbar \omega \hat{b}'_j(\omega) \hat{b}'^\dagger_j(\omega), \\
    \hat{\mathcal{H}}'_\mathrm{Ph1-Ph2} & = \sum^{N'_\mathrm{ph}}_{j=1} \int_{-\infty}^\infty \dd{\omega} \hbar\sqrt{\frac{\kappa'_{\mathrm{ph}, j}}{2\pi}} \left[ \hat{b}'^\dagger_j(\omega) \hat{a}'_j + \hat{b}'_j(\omega) \hat{a}'^\dagger_j \right], \\
    \hat{\mathcal{H}}'_\mathrm{M-Ph} &= \sum_{\alpha, j} \hbar \Omega'_{\alpha j} \hat{\bar{\mu}}_\alpha \left( \hat{a}'^\dagger_j + \hat{a}'_j \right).
\end{align}
$N'_\mathrm{ph}$, $\omega'_{\mathrm{ph}, j}$, $ \Omega'_{\alpha j}$, $\kappa'_{\mathrm{ph}, j}$, $\hat{a}'^\dagger_j$ and $\hat{b}'_j(\omega)$ follow the similar definition as their counterparts in dissipative CQED-DDI. The only difference between $\hat{\mathcal{H}}'_\mathrm{T}$ and $\hat{\mathcal{H}}_\mathrm{T}$ is that the photonic modes in $\hat{\mathcal{H}}'_\mathrm{T}$ are no longer separated as the free-space and the scattering parts; instead, they are treated as a whole. Then, we again separate the effective microscopic model into the system $\hat{\mathcal{H}}'_\mathrm{S}$, the bath $\hat{\mathcal{H}}'_\mathrm{B}$ and the system-bath interaction $\hat{\mathcal{H}}'_\mathrm{S-B}$ as
\begin{align}
    \hat{\mathcal{H}}'_\mathrm{T} = \hat{\mathcal{H}}'_\mathrm{S} + \hat{\mathcal{H}}'_\mathrm{B} + \hat{\mathcal{H}}'_\mathrm{S-B}.
\end{align}
The system Hamiltonian includes the molecules $\hat{H}_\mathrm{M}$, the discrete primary photonic modes $\hat{\mathcal{H}}'_\mathrm{Ph1}$ and the interactions between them $\hat{\mathcal{H}}'_\mathrm{M-Ph}$, i.e.,
\begin{align}
    \hat{\mathcal{H}}'_\mathrm{S} & = \hat{H}_\mathrm{M} + \hat{\mathcal{H}}'_\mathrm{Ph1} + \hat{\mathcal{H}}'_\mathrm{M-Ph};
\end{align}
the bath Hamiltonian includes the continuous secondary photonic modes $\hat{\mathcal{H}}'_\mathrm{Ph2}$, i.e.,
\begin{align}
    \hat{\mathcal{H}}'_\mathrm{B} & = \hat{\mathcal{H}}'_\mathrm{Ph2};
\end{align}
the system-bath interaction Hamiltonian includes the interactions between the discrete primary photonic modes and the continuous secondary photonic modes $\hat{\mathcal{H}}'_\mathrm{Ph1-Ph2}$, i.e.,
\begin{align}
\hat{\mathcal{H}}'_\mathrm{S-B} & = \hat{\mathcal{H}}'_\mathrm{Ph1-Ph2}.
\end{align}
Finally, by considering the initial condition that the bath modes (the photonic modes) are in the vacuum state and tracing out the degrees of freedom of the bath, we can obtain dissipative CQED as
\begin{align}
\nonumber
    & \pdv{t} \hat{\rho}'_\mathrm{S}(t) =  -\frac{i}{\hbar} \left[ \hat{H}_\mathrm{M} + \hat{\mathcal{H}}'_\mathrm{Ph1} + \hat{\mathcal{H}}'_\mathrm{M-Ph}, \hat{\rho}'_\mathrm{S}(t) \right] \\
    & \quad  + \sum_j \kappa'_{\mathrm{ph}, j} \left[ \hat{a}'_j \hat{\rho}'_\mathrm{S}(t) \hat{a}'^\dagger_j - \frac{1}{2} \left\{ \hat{a}'^\dagger_j \hat{a}'_j, \hat{\rho}'_\mathrm{S}(t) \right\} \right].
\label{Eq:EOM_rho_CQED_woDDI}
\end{align}

Note that Eq.~(\ref{Eq:EOM_rho_CQED_woDDI}) is the main result of dissipative CQED. In addition, the generalized spectral density $\overline{\overline{\mathfrak{J}}}\vphantom{a}'(\omega)$ of the effective microscopic model $\hat{\mathcal{H}}'_\mathrm{T}$ is now expressed as
\begin{align}
    \overline{\overline{\mathfrak{J}}}\vphantom{a}'(\omega) = \frac{1}{\pi} \mathrm{Im} \left[  \overline{\overline{\boldsymbol{\Omega}}}\vphantom{a}' \cdot   (\overline{\overline{\mathcal{H}}}\vphantom{a}'_\mathrm{eff} - \omega  \overline{\overline{\mathbf{I}}}_{N'_\mathrm{ph}})^{-1} \cdot  (\overline{\overline{\boldsymbol{\Omega}}}\vphantom{a}')^\mathrm{T} \right],
\label{Eq:J_CQED}
\end{align}
where $\mathcal{H}\vphantom{a}'_{\mathrm{eff},jk} = (\omega'_{\mathrm{ph}, j} - \frac{i}{2} \kappa'_{\mathrm{ph}, j}) \delta_{jk}$. By fitting the total generalized spectral density $\overline{\overline{\mathbf{J}}}(\omega)$ to $\overline{\overline{\mathfrak{J}}}\vphantom{a}'(\omega)$, we can obtain the parameters $\omega'_{\mathrm{ph}, j}$, $ \Omega'_{\alpha j}$ and $\kappa'_{\mathrm{ph}, j}$ in dissipative CQED. 

However, expanding $\overline{\overline{\mathfrak{J}}}\vphantom{a}'(\omega)$ in Eq.~(\ref{Eq:J_CQED}), we find that its matrix element is indeed a sum of Lorentzians, i.e.,
\begin{align}
    \mathfrak{J}'_{\alpha \beta}(\omega) = \sum_{j=1}^{N'_\mathrm{ph}} \frac{\Omega'_{\alpha j} \Omega'_{\beta j} }{\pi} \frac{\kappa'_j/2}{(\omega - \omega'_{\mathrm{ph},j})^2 + (\kappa'_j/2)^2}. 
\end{align}
Obviously, the free-space generalized spectral density, i.e., 
\begin{align}
\nonumber
    J_{0, \alpha \beta}(\omega) & = \frac{\omega^2}{\pi \hbar \epsilon_0 c^2} \boldsymbol{\mu}_\alpha \cdot \mathrm{Im} \overline{\overline{\mathbf{G}}}_0(\mathbf{r}_\alpha, \mathbf{r}_\beta,\omega) \cdot \boldsymbol{\mu}_\beta \\
\nonumber
    & = \frac{k_0^3}{4 \pi^2 \hbar \varepsilon_0} \biggl\{ \left[ \boldsymbol{\mu}_\alpha \cdot \boldsymbol{\mu}_\beta - \left( \boldsymbol{\mu}_\alpha \cdot \mathbf{n}_R \right) \left( \boldsymbol{\mu}_\beta \cdot \mathbf{n}_R \right) \right] \\
\nonumber    
    & \quad \times \frac{\sin{(k_0 R)}}{k_0 R} + \left[ \boldsymbol{\mu}_\alpha \cdot \boldsymbol{\mu}_\beta - 3 \left( \boldsymbol{\mu}_\alpha \cdot \mathbf{n}_R \right) \left( \boldsymbol{\mu}_\beta \cdot \mathbf{n}_R \right) \right] \\
    & \quad \times \left[ \frac{\cos{(k_0 R)}}{(k_0 R)^2} - \frac{\sin{(k_0 R)}}{(k_0 R)^3} \right] \biggr\}
\end{align}
($k_0 = \omega / c$ and $ \mathbf{r}_\alpha - \mathbf{r}_\beta \equiv R \mathbf{n}_R$), cannot be expressed in terms of a limited number of Lorentzians, which implies that the free-space effects generally cannot be entirely captured using a limited number of dissipative photonic modes. For instance, the free-space dipole-dipole interactions are the combined effects of the whole electromagnetic spectrum, including infinite photonic modes that are off-resonant with the molecules. As a result, dissipative CQED may fail to accurately describe light-matter interactions in the exciton-polariton systems involving multiple molecules.

Contrary to dissipative CQED, dissipative CQED-DDI offers an advantage because it can fully encapsulate the free-space effects. The cleverness of dissipative CQED-DDI lies in the fact that the free-space and the dielectric-induced effects are treated separately. For the free-space effects, we consider the entire free-space electromagnetic spectrum for the free-space photonic modes. Then, taking into account the Markovian nature of the free-space electromagnetic environment, we can regard it as a bath and simply trace it out, resulting in free-space spontaneous emissions, dephasings and dipole-dipole interactions (the Lamb shifts are discarded for convenience). The dielectric-induced effects, on the other hand, can be primarily included by considering a particular range of the electromagnetic spectrum, as the medium shapes the electromagnetic spectrum only within a certain frequency range and exhibits transparency in the high-frequency domain. Therefore, the dielectric-induced effects can be effectively modeled by molecules coupled with a few dissipative photonic modes.

We summarize the comparison between MQED, dissipative CQED-DDI and dissipative CQED in FIG.~\ref{Fig2}. Here we would like to emphasize that the main distinction between dissipative CQED-DDI and dissipative CQED is rooted in how the free-space effects are treated. 

\section{Numerical Demonstration and Discussion}
\label{Sec:Numerical Demonstration and Discussion}

In the previous sections, we have derived dissipative CQED-DDI and discussed its difference from dissipative CQED. In this section, we numerically demonstrate the validity and advantage of dissipative CQED-DDI by applying this method to study the excited-state population of molecules above a plasmonic surface and comparing the results to those obtained from several other methods, including dissipative CQED, MQED wavefunction approach \cite{Chuang2023} (MQED-WF) and MQED density matrix approach under the Markov approximation \cite{Dung2002_2} (MQED-DMMA). We summarize the main features of MQED-WF and MQED-DMMA in the following:

(i) MQED wavefunction approach (MQED-WF):
\begin{align}
\nonumber
    & \dv{t} C^{\mathrm{E_{\alpha}},\left\{0\right\}}(t) = \\
\nonumber
    & \quad - \sum_\beta \int_0^t \dd{t'} \int_0^\infty \dd{\omega} J_{\alpha \beta}(\omega) e^{-i \left( \omega - \omega_\mathrm{M} \right) (t-t')} C^{\mathrm{E_{\beta}},\left\{0\right\}}(t') \\
\nonumber
    & \quad - \sum_{\beta\neq\alpha} \int_0^t \dd{t'} \int_0^\infty \dd{\omega} J_{\beta \beta}(\omega) e^{-i \left( \omega + \omega_\mathrm{M} \right) (t+t')}  C^{\mathrm{E_{\alpha}},\left\{0\right\}}(t') \\
    & \quad - \sum_{\beta\neq\alpha} \int_0^t \dd{t'} \int_0^\infty \dd{\omega} J_{\beta \alpha}(\omega) e^{-i \left( \omega + \omega_\mathrm{M} \right) (t+t')}  C^{\mathrm{E_{\beta}},\left\{0\right\}}(t'),
\label{Eq:MQED_WF}
\end{align}
where $C^{\mathrm{E_{\alpha}},\left\{0\right\}}(t)$ is the probability amplitude for the state that $\alpha$ molecule is in its electronically excited state with zero polariton. The population of $\alpha$ in its excited state is given by $P^{\mathrm{E_{\alpha}},\left\{0\right\}}(t) = \abs{C^{\mathrm{E_{\alpha}},\left\{0\right\}}(t)}^2$.

(ii) MQED density matrix approach under the Markov approximation (MQED-DMMA):
\begin{align}
\nonumber
    & \pdv{t} \hat{\rho}_\mathrm{M}(t) =  -\frac{i}{\hbar} \left[ \hat{H}_\mathrm{M} + \hat{\mathcal{H}}^\mathrm{Sc}_\mathrm{LS} + \hat{\mathcal{H}}_\mathrm{DDI}, \hat{\rho}_\mathrm{M}(t) \right] \\ 
    & \quad + \sum_{\alpha, \beta} \tilde{\Gamma}_{\alpha \beta} \left[ \hat{\sigma}^{(-)}_\beta \hat{\rho}_\mathrm{M}(t) \hat{\sigma}^{(+)}_\alpha - \frac{1}{2} \left\{ \hat{\sigma}^{(+)}_\alpha \hat{\sigma}^{(-)}_\beta, \hat{\rho}_\mathrm{M}(t) \right\} \right],
\label{Eq:MQED_DMMA}
\end{align}
with
\begin{align}
    \hat{\mathcal{H}}^\mathrm{Sc}_\mathrm{LS} & = \sum_{\alpha} \Delta^\mathrm{Sc}_\alpha \hat{\sigma}^{(+)}_\alpha \hat{\sigma}^{(-)}_\alpha, \\
    \hat{\mathcal{H}}_\mathrm{DDI} & = \sum_{\substack{\alpha,\beta\\
    \alpha\neq\beta}} \tilde{V}_{\alpha \beta} \hat{\sigma}^{(+)}_\alpha \hat{\sigma}^{(-)}_\beta,
\end{align}
where $\hat{\rho}_\mathrm{M}(t)$ is the reduced density matrix of the molecules. $\tilde{\Gamma}_{\alpha \beta}$ and $\tilde{V}_{\alpha \beta}$ are obtained by substituting $\overline{\overline{\mathbf{G}}}_0(\mathbf{r}_\alpha,\mathbf{r}_\beta,\omega)$ with $\overline{\overline{\mathbf{G}}}(\mathbf{r}_\alpha,\mathbf{r}_\beta,\omega)$ in the expression of $\tilde{\Gamma}^0_{\alpha \beta}$ and $\tilde{V}^0_{\alpha \beta}$, respectively, and $\Delta^\mathrm{Sc}_\alpha$ is obtained by substituting $\overline{\overline{\mathbf{G}}}_0(\mathbf{r}_\alpha,\mathbf{r}_\beta,\omega)$ with $\overline{\overline{\mathbf{G}}}_\mathrm{Sc}(\mathbf{r}_\alpha,\mathbf{r}_\beta,\omega)$ in the expression of $\Delta^0_\alpha$. The population of $\alpha$ in its excited state is given by $P^{\mathrm{E_{\alpha}},\left\{0\right\}}(t) = \mathrm{Tr}\left[ \hat{\sigma}^{(+)}_\alpha \hat{\sigma}^{(-)}_\alpha \hat{\rho}_\mathrm{M}(t) \right]$. Note that in Eq.~(\ref{Eq:MQED_DMMA}), we have dropped the free-space Lamb shift. In addition, since this approach is derived under the Markov approximation, it is only valid in the weak light-matter interaction regime.

It should be noted that for dissipative CQED-DDI and dissipative CQED, the non-zero value of $\overline{\overline{\mathfrak{J}}}_\mathrm{Sc}(\omega)$ and $\overline{\overline{\mathfrak{J}}}\vphantom{a}'(\omega)$ for $\omega < 0$ can induce artificial pumping to the system when the counter-rotating interactions exist, yielding inaccurate quantum dynamics. To resolve the problem, we further implement the rotating-wave approximation, e.g., $ \hat{\mathcal{H}}_\mathrm{int1} \approx \sum_{\alpha, j} \hbar \Omega_{\alpha j} \left( \hat{\sigma}^{(-)}_\alpha \hat{a}^\dagger_j + \hat{\sigma}^{(+)}_\alpha \hat{a}_j \right)$, in dissipative CQED-DDI and dissipative CQED, considering that the exciton-polariton interaction strengths are not too strong in our cases of studies. However, it is crucial to acknowledge that when the interaction strengths are too strong, such as in the ultra-strong and deep-strong interaction regimes, making the rotating-wave approximation can also lead to incorrect dynamics. In such scenarios, the artificial pumping issue can be alternatively mitigated by formulating the dissipators in terms of the dressed basis \cite{Beaudoin2011,Settineri2018} or by suppressing the negative component of the generalized spectral density utilizing the interacting-mode model \cite{Lednev2023}. 

\begin{figure}[t]
    \centering
    \includegraphics[width=0.45\textwidth]{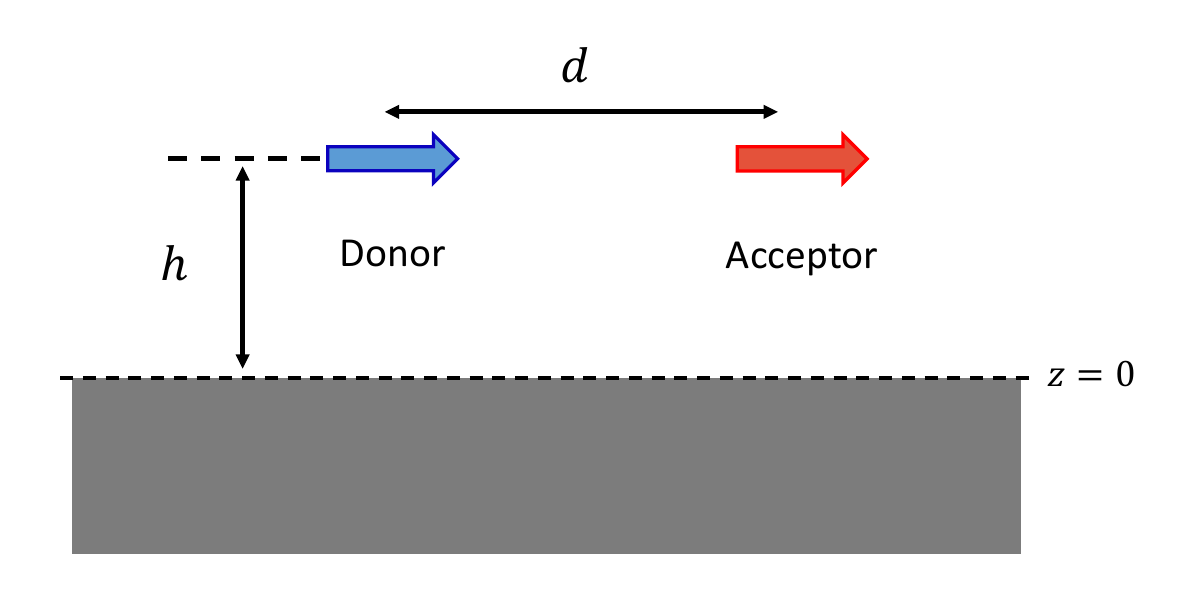}
    \caption{Schematic illustration of two molecules (a donor and an acceptor) above a plasmonic surface. The molecule-surface distance and intermolecular distance are denoted as $h$ and $d$, respectively.}
    \label{Fig3}
\end{figure}

For MQED-WF and MQED-DMMA, we do not apply the rotating-wave approximation, so the effects of counter-rotating interactions are incorporated. In the following study, population dynamics calculated using MQED-WF will serve as the reference for examining the results obtained via other methods. On the other hand, since MQED-DMMA is only valid in the weak-coupling regime, it will serve as a criterion for us to verify whether the light-matter interaction strength is strong or weak, e.g., the consistency between MQED-DMMA and MQED-WF indicates that the light-matter interaction strength is weak.

The system in our numerical study comprises a pair of molecules (or a single molecule) above a plasmonic surface, as depicted in FIG.~\ref{Fig3}. The surface is modeled by the dielectric function:
\begin{align}
    \varepsilon_\mathrm{r}(\mathbf{r},\omega) = 
    \begin{cases}
        1,    & z>0, \\
        1 - 5^2/(\omega^2+0.1i\omega),    & z<0. 
    \end{cases} 
    \label{Eq:Dielectrics}
\end{align}
The transition frequency and transition dipole moment of the molecules are designated as $\hbar\omega_\mathrm{D} = \hbar\omega_\mathrm{A} \equiv \hbar \omega_\mathrm{M} = 3.525$ eV and $\abs{\boldsymbol{\mu}_\mathrm{D}}=\abs{\boldsymbol{\mu}_\mathrm{A}}=10$ Debye, respectively. Note that in this simple system, we can systematically modify exciton-polariton interaction strengths and free-space intermolecular dipole-dipole interactions through varying molecule-metal distance $h$
and donor-acceptor distance $d$.

\begin{figure*}[!t]
    \centering
    \includegraphics[width=0.8\textwidth]{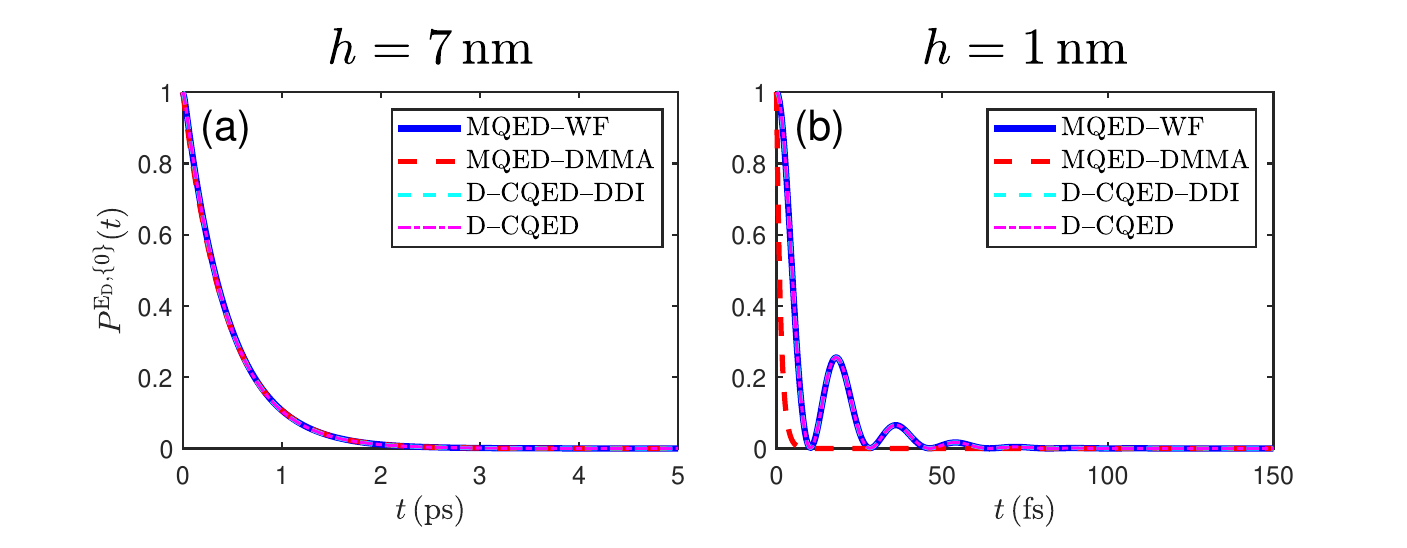}
    \caption{Population dynamics of a single molecule above a plasmonic surface for (a) $h=7$ nm and (b) $h=1$ nm. The population dynamics is obtained via four methods: (1) MQED-WF [recall Eq.~(\ref{Eq:MQED_WF})]; (2) MQED-DMMA [recall Eq.~(\ref{Eq:MQED_DMMA})]; (3) D-CQED-DDI (dissipative CQED-DDI) [recall Eq.~(\ref{Eq:EOM_rho2_CQED})]; D-CQED (dissipative CQED) [recall Eq.~(\ref{Eq:EOM_rho_CQED_woDDI})].}
    \label{Fig4}
\end{figure*}

\subsection{Quantum dynamics of a single molecule: Absence of free-space dipole-dipole interaction}

Exciton-polariton dynamics for a single molecule is a special case in which  intermolecular dipole-dipole interactions do not exist. To examine whether dissipative CQED-DDI and dissipative CQED can capture the main phenomena from weak to strong light-matter interactions, we compare the population dynamics obtained from dissipative CQED-DDI and dissipative CQED with those calculated via MQED-WF and MQED-DMMA.
We consider two different cases, including $h=7$ nm and $h=1$ nm. In addition, in both cases, we use two modes, i.e., $N_\mathrm{ph}~ (N'_\mathrm{ph}) = 2$, to fit the scattering generalized spectral density for dissipative CQED-DDI and the generalized spectral density for dissipative CQED. The fitting parameters are shown in Appendix \ref{Appendix:Parameters}, and the calculated excited-state population dynamics of the donor $P^{\mathrm{E_D},{0}}(t)$ are shown in FIG.~\ref{Fig4}.

\begin{figure*}[!t]
    \centering
    \includegraphics[width=1\textwidth]{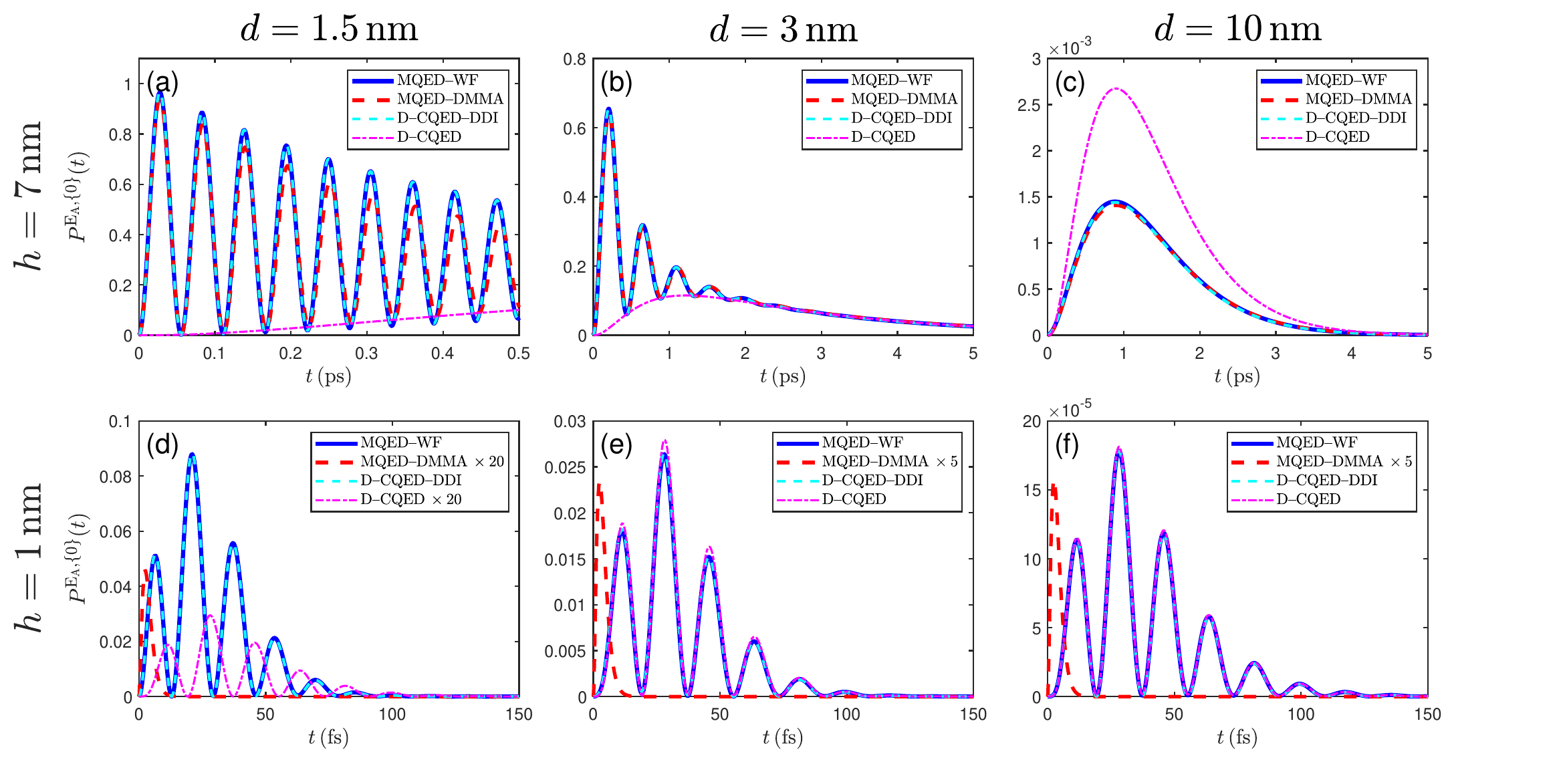}
    \caption{Population dynamics of the acceptor for (a) $h=7$ nm and $d=1.5$ nm, (b) $h=7$ nm and $d=3$nm, (c) $h=7$ nm and $d=10$nm, (d) $h=1$ nm and $d=1.5$, (e) $h=1$ nm and $d=3$ nm, and (f) $h=1$ nm and $d=10$ nm. The population dynamics is obtained via four methods: (1) MQED-WF [recall Eq.~(\ref{Eq:MQED_WF})]; (2) MQED-DMMA [recall Eq.~(\ref{Eq:MQED_DMMA})]; (3) D-CQED-DDI (dissipative CQED-DDI) [recall Eq.~(\ref{Eq:EOM_rho2_CQED})]; D-CQED (dissipative CQED) [recall Eq.~(\ref{Eq:EOM_rho_CQED_woDDI})].}
    \label{Fig5}
\end{figure*}

To better understand the exciton-polariton dynamics, first, we compare the population dynamics calculated by MQED-WF and MQED-DMMA and then verify whether the exciton-polariton interaction is strong or weak. In FIG.~\ref{Fig4}(a), when $h=7$ nm, there is a match between $P^{\mathrm{E_D},{0}}(t)$ obtained using MQED-WF and MQED-DMMA. Given that MQED-DMMA is derived under the Markov approximation, the consistency between MQED-DMMA and MQED-WF suggests that for this certain scenario of $h=7$ nm, we are within the single-molecule weak-coupling regime. Conversely, in FIG.~\ref{Fig4}(b), when $h=1$ nm, the population $P^{\mathrm{E_D},{0}}(t)$ calculated by MQED-DMMA diverges from that calculated by MQED-WF, indicating a transition to the single-molecule strong-coupling regime. Second, we compare the population dynamics calculated by MQED-WF, dissipative CQED-DDI and dissipative CQED to examine the validity of dissipative CQED-DDI and dissipative CQED. In both scenarios, with $h=7$ nm in FIG.~\ref{Fig4}(a) and $h=1$ nm in FIG.~\ref{Fig4}(b), the populations $P^{\mathrm{E_D},{0}}(t)$ calculated by both dissipative CQED-DDI and dissipative CQED align with those calculated by MQED-WF. This suggests that in the presence of a single molecule (when dipole-dipole interactions are absent), both dissipative CQED-DDI and dissipative CQED yield precise population dynamics. 

To summarize, in the case of a single molecule, both dissipative CQED-DDI and dissipative CQED can yield the same quantum dynamics, no matter in the weak or strong coupling conditions.

\subsection{Quantum dynamics of a pair of molecules: Presence of free-space dipole-dipole interaction}

In this section, we move on to the case of two molecules. We choose the intermolecular distance $d$ to be 1.5 nm, 3 nm, and 10 nm and study the population dynamics of the acceptor $P^{\mathrm{E_A},{0}}(t)$. In each case, We use four modes, i.e., $N_\mathrm{ph}~ (N'_\mathrm{ph}) = 4$, to fit the scattering generalized spectral density for dissipative CQED-DDI and the generalized spectral density for dissipative CQED. The fitting parameters are shown in Appendix \ref{Appendix:Parameters}. In addition, the initial condition is set to be $P^{\mathrm{E_\alpha},{0}}(t=0) (C^{\mathrm{E_\alpha},{0}}(t=0)) = \delta_{\alpha \mathrm{D}}$. The calculated excited-state population dynamics of the acceptor $P^{\mathrm{E_A},{0}}(t)$ are plotted in FIG.~\ref{Fig5}.

As in the single-molecule case, we first compare the population dynamics calculated by MQED-WF and MQED-DMMA and verify the weak/strong coupling regime. The results are similar to the single-molecule case. In FIGs.~\ref{Fig5}(a) to \ref{Fig5}(c), when $h=7$ nm, which is the single-molecule weak-coupling regime, the populations $P^{\mathrm{E_A},{0}}(t)$ calculated via MQED-DMMA almost match those calculated via MQED-WF, except for slight deviation in the condition of $d=1.5$ nm in FIG.~\ref{Fig5}(a). The slight deviation between the blue line and the red dashed line in FIG.~\ref{Fig5}(a) indicates that the population dynamics is no longer Markovian, which is likely due to the strong free-space dipole-dipole interaction between the donor and the acceptor that are close to each other. Nevertheless, MQED-DMMA can still capture the oscillatory behavior of $P^{\mathrm{E_A},{0}}(t)$. On the contrary, in FIGs.~\ref{Fig5}(d) to \ref{Fig5}(f), when $h=1$ nm, the populations $P^{\mathrm{E_A},{0}}(t)$ calculated via MQED-DMMA significantly deviate from those calculated via MQED-WF because in the strong-coupling regime the Markov approximation fails and MQED-DMMA cannot capture the correct population dynamics.


Next, we turn our attention to the comparison between MQED-WF, dissipative CQED-DDI and dissipative CQED to examine the validity of dissipative CQED-DDI and dissipative CQED. In FIGs.~\ref{Fig5}(a) to \ref{Fig5}(f), $P^{\mathrm{E_A},{0}}(t)$ obtained from dissipative CQED-DDI align perfectly with those obtained from MQED-WF. The consistency between these two methods supports the validity of dissipative CQED-DDI from small to large molecule-metal distance (weak to strong exciton-polariton coupling strength) and from small to large intermolecular distance (weak to strong free-space dipole-dipole interaction). On the contrary, $P^{\mathrm{E_A},{0}}(t)$ obtained from dissipative CQED only match those obtained from MQED-WF in the cases when $h=1$ nm and $d=3$ nm in FIG.~\ref{Fig5}(e) and $h=1$ nm and $d=10$ nm in FIG.~\ref{Fig5}(f). The result suggests that dissipative CQED is only applicable when the molecules are in proximity to the surface (strong exciton-polariton coupling) but are not close to each other (weak free-space dipole-dipole interaction). The comparison among the four models (or methods) indicates that direct intermolecular interactions, e.g., free-space dipole-dipole interactions, generally are a non-negligible component in CQED-like models (which consider only a single or a few photonic modes) when applied to the exciton-polariton systems that involve more than a single molecule.

To summarize, we have shown that in a multi-molecule system, dissipative CQED-DDI offers a versatile tool for studying the quantum dynamics of molecules in various configurations, including conditions spanning from weak to strong exciton-polariton coupling strengths and free-space dipole-dipole interactions. In contrast, dissipative CQED is restricted to the strong exciton-polariton coupling strength with weak free-space dipole-dipole interaction. Our result provides an important insight into exciton-polariton formation involving multiple molecules and points out the importance of direct intermolecule interactions in the (dissipative) CQED model.

\section{Conclusions}
\label{Sec:Conclusion}

In this study, 
we have analytically and numerically shown that "\textit{one should include direct intermolecular interactions in the CQED model
when studying exciton-polariton systems involving multiple molecules}". Analytically, we derived a dissipative CQED model including free-space dipole-dipole interactions (CQED-DDI) from an effective microscopic Hamiltonian based on MQED. Dissipative CQED-DDI successfully encapsulates the influence of light-matter interactions in dielectric environments and effectively separates them into the free-space effects and the dielectric-induced effects. The free-space effects include spontaneous emissions, dephasings and dipole-dipole interactions in the free-space; the dielectric-induced effects include exciton-polariton(photon) interactions and photonic losses. From a theoretical point of view, in comparison with dissipative CQED (which does not include any direct intermolecular interactions), dissipative CQED-DDI adequately takes into account the free-space effects that originate from the full electromagnetic spectrum. 

In addition, we numerically demonstrate the validity and advantage of dissipative CQED-DDI by applying this method to study the exciton-polariton dynamics (excited-state population dynamics of molecules above a plasmonic surface) and comparing the result with three different methods, including (i) MQED wavefunction approach (MQED-WF), (ii) MQED density matrix approach under the Markov approximation (MQED-DMMA) and (iii) dissipative CQED. MQED-WF serves as a reference for examining the dynamics obtained from dissipative CQED-DDI, and MQED-DMMA provides a simple standard for us to verify whether the light-matter interaction strength is strong or weak. In addition, the comparison between dissipative CQED-DDI and dissipative CQED can answer the question of whether one should include the free-space dipole-dipole interactions in the (dissipative) CQED model. In the case of a single molecule, both dissipative CQED-DDI and dissipative CQED yield accurate population dynamics regardless of the exciton-polariton coupling strength (the molecule-surface distance). In the case of multiple molecules, dissipative CQED-DDI continues to provide precise dynamics from weak to strong exciton-polariton coupling strengths (large molecule-surface distance) and from weak to strong free-space dipole-dipole interactions (small to large intermolecular distance). On the contrary, dissipative CQED-DDI performs well only when the exciton-polariton coupling strength is strong (small molecule-surface distance) and the free-space dipole-dipole interaction us weak (large intermolecular distance). Our result indicates that the direct intermolecular interactions included in dissipative CQED-DDI play a key role when studying the exciton-polariton systems involving multiple molecules.  Considering that many of the studies that adopt the CQED model (or its extensions) do not consider direct intermolecular interactions, we believe that this work can provide an important insight into light-matter interaction and polariton chemistry.


\begin{acknowledgments}
Hsu thanks Academia Sinica (AS-CDA-111-M02), National Science and Technology Council (111-2113-M-001-027-MY4) and Physics Division, National Center for Theoretical Sciences (112-2124-M-002-003) for the financial support.
\end{acknowledgments}

\section*{data availability}
The data that support the findings of this study are available from the corresponding author upon reasonable request.

\appendix

\section{Derivation of Eq.~(\ref{Eq:EOM_rho_CQED})}
\label{Sec:AppendixA}

To derive Eq.~(\ref{Eq:EOM_rho_CQED}), we begin with the system-bath Hamiltonian in Eq.~(\ref{Eq:Hamiltonian_SB}). In the Heisenberg picture, the time-dependent dynamics of an operator follows
\begin{align}
    \pdv{t} \hat{O}(t) = \frac{i}{\hbar} \left[ \hat{\mathcal{H}}_\mathrm{T}(t), \hat{O}(t) \right].
\label{Eq:Heisenberg}
\end{align}
For a system operator $\hat{O}_\mathrm{S}$, e.g., $\hat{\sigma}^{(+)}_\alpha$ or $\hat{a}_j$, that commutes with the bath operators, its equation of motion can be expressed as
\begin{align}
    \pdv{t} \hat{O}_\mathrm{S}(t) = \frac{i}{\hbar} \left[ \hat{\mathcal{H}}_\mathrm{S}(t) +  \hat{\mathcal{H}}_\mathrm{S-B}(t), \hat{O}_\mathrm{S}(t) \right].
\label{Eq:Commutator_OS}
\end{align}
The commutator $\frac{i}{\hbar} \left[ \hat{\mathcal{H}}_\mathrm{S-B}(t), \hat{O}(t) \right]$ can be further decomposed as
\begin{align}
\nonumber
    \frac{i}{\hbar} \left[ \hat{\mathcal{H}}_\mathrm{S-B}(t), \hat{O}(t) \right] & = \frac{i}{\hbar} \left[ \hat{\mathcal{H}}^0_\mathrm{M-Ph}(t), \hat{O}_\mathrm{S}(t) \right] \\ 
    & \quad + \frac{i}{\hbar} \left[ \hat{\mathcal{H}}_\mathrm{Ph1-Ph2}(t), \hat{O}_\mathrm{S}(t) \right].
\label{Eq:Commutator_SB}
\end{align}
The first term on the right-hand side of Eq.~(\ref{Eq:Commutator_SB}) can be expanded as
\begin{widetext}
\begin{align}
    \frac{i}{\hbar} \left[ \hat{\mathcal{H}}^0_\mathrm{M-Ph}(t), \hat{O}_\mathrm{S}(t) \right] = i \sum_{\alpha, l} \int_{-\infty}^{\infty} \dd{\omega} \hbar g_{\alpha l}(\omega) \left\{ \hat{c}^\dagger_l(\omega,t) \left[ \hat{\bar{\mu}}_\alpha(t), \hat{O}_\mathrm{S}(t) \right] + \left[ \hat{\bar{\mu}}_\alpha(t), \hat{O}_\mathrm{S}(t) \right] \hat{c}_l(\omega,t) \right\}.
\label{Eq:Commutator_MPh0}
\end{align}
The equations of motion of $\hat{c}_l(\omega,t)$ and $\hat{c}^\dagger_l(\omega,t)$ also follow the Heisenberg equation in Eq.~(\ref{Eq:Heisenberg}), resulting in
\begin{subequations}
\begin{gather}
    \pdv{t} \hat{c}_l(\omega,t) = -i\omega \hat{c}_l(\omega,t) - i \sum_\alpha g_{\alpha l}(\omega) \hat{\bar{\mu}}_\alpha(t), 
\label{Eq:dc_dynamics} \\
    \pdv{t} \hat{c}^\dagger_l(\omega,t) = i\omega \hat{c}^\dagger_l(\omega,t) + i \sum_\alpha g_{\alpha l}(\omega) \hat{\bar{\mu}}_\alpha(t).
\label{Eq:dc_dagger_dynamics}
\end{gather}
\end{subequations}
Formally integrating Eqs.~(\ref{Eq:dc_dynamics}) and (\ref{Eq:dc_dagger_dynamics}), we obtain
\begin{subequations}
\begin{gather}
    \hat{c}_l(\omega,t) = \hat{c}_{l, \mathrm{free}}(\omega,t) - i \sum_\alpha \int_0^{t} \dd{t'} e^{-i\omega (t-t')} g_{\alpha l}(\omega) \hat{\bar{\mu}}_\alpha(t'), 
\label{Eq:c_dynamics} \\
    \hat{c}^\dagger_l(\omega,t) = \hat{c}^\dagger_{l, \mathrm{free}}(\omega,t) + i \sum_\alpha \int_0^{t} \dd{t'} e^{i\omega (t-t')} g_{\alpha l}(\omega) \hat{\bar{\mu}}_\alpha(t'),
\label{Eq:c_dagger_dynamics}
\end{gather}
\end{subequations}
where $\hat{c}^\dagger_{l, \mathrm{free}}(\omega,t) = e^{i \omega t} \hat{c}^\dagger_l(\omega,0)$ and $\hat{c}_{l, \mathrm{free}}(\omega,t) = e^{-i \omega t} \hat{c}_l(\omega,0)$ are the free-evolution of $\hat{c}^\dagger_l(\omega,t)$ and $\hat{c}_l(\omega,t)$.
Plugging Eqs.~(\ref{Eq:c_dynamics}) and (\ref{Eq:c_dagger_dynamics}) into Eq.~(\ref{Eq:Commutator_MPh0}), the right hand side becomes
\begin{align}
\nonumber
    & i \sum_{\alpha, l} \int_{-\infty}^{\infty} \dd{\omega} g_{\alpha l}(\omega) \left\{ \left[ \hat{c}^\dagger_{l, \mathrm{free}}(\omega,t) + i \sum_\beta \int_0^{t} \dd{t'} e^{i\omega (t-t')} g_{\beta l}(\omega) \hat{\bar{\mu}}_\beta(t')\right] \left[ \hat{\bar{\mu}}_\alpha(t), \hat{O}_\mathrm{S}(t) \right] \right. \\
\nonumber
    & \quad \quad \left.+ \left[ \hat{\bar{\mu}}_\alpha(t), \hat{O}_\mathrm{S}(t) \right] \left[ \hat{c}_{l, \mathrm{free}}(\omega,t) - i \sum_\beta \int_0^{t} \dd{t'} e^{-i\omega (t-t')} g_{\beta l}(\omega) \hat{\bar{\mu}}_\beta(t') \right] \right\} \\
    & \quad = - \sum_{\alpha, \beta, l} \int_{-\infty}^{\infty} \dd{\omega} \int_0^{t} \dd{t'} g_{\alpha l}(\omega) g_{\beta l}(\omega) \left\{  e^{i\omega (t-t')}  \hat{\bar{\mu}}_\beta(t') \left[ \hat{\bar{\mu}}_\alpha(t), \hat{O}_\mathrm{S}(t) \right] - e^{-i\omega (t-t')}  \left[ \hat{\bar{\mu}}_\alpha(t), \hat{O}_\mathrm{S}(t) \right] \hat{\bar{\mu}}_\beta(t') \right\} + \hat{C}_\mathrm{free}(t),
\label{Eq:Commutator_MPh0_2}
\end{align}
where
\begin{align}
    \hat{C}_\mathrm{free}(t) = i \sum_{\alpha, l} \int_{-\infty}^{\infty} \dd{\omega} g_{\alpha l}(\omega) \left\{   \hat{c}^\dagger_{l, \mathrm{free}}(\omega,t) \left[ \hat{\bar{\mu}}_\alpha(t), \hat{O}_\mathrm{S}(t) \right] + \left[ \hat{\bar{\mu}}_\alpha(t), \hat{O}_\mathrm{S}(t) \right] \hat{c}_{l, \mathrm{free}}(\omega,t) \right\}. 
\end{align}
According to Eqs.~(\ref{Eq:g_MPh0}) to (\ref{Eq:S}), we have the identity
\begin{align}
    \sum_l g_{\alpha l}(\omega) g_{\beta l}(\omega) = \theta(\omega) J_{0, \alpha \beta} (\omega).
\end{align}
In addition, considering that the interactions between molecules and the free-space fields are typically weak, the memory effect can be neglected, and we can apply the Markov approximation  to the time-dependent integral, i.e., we make the substitutions
\begin{gather}
\nonumber
    \int_0^t \dd{t'} \rightarrow \int_{-\infty}^t \dd{t'}, \\
\nonumber
    \hat{\bar{\mu}}_\alpha(t') \rightarrow e^{i \omega_\alpha (t'-t)} \hat{\sigma}^{(+)}_\alpha(t) + e^{-i \omega_\alpha (t'-t)} \hat{\sigma}^{(-)}_\alpha(t).
\end{gather}
Therefore, Eq.~(\ref{Eq:Commutator_MPh0_2}) can be transformed to
\begin{align}
\nonumber
    & - \sum_{\alpha, \beta} \int_0^{\infty} \dd{\omega} \int_{-\infty}^{t} \dd{t'} J_{0, \alpha \beta}(\omega) \left\{  \left[ e^{i (\omega - \omega_\beta) (t-t')}  \hat{\sigma}^{(+)}_\beta(t) + e^{i (\omega + \omega_\beta) (t-t')}  \hat{\sigma}^{(-)}_\beta(t) \right] \left[ \hat{\bar{\mu}}_\alpha(t), \hat{O}_\mathrm{S}(t) \right] \right. \\
    & \quad \quad \left. -  \left[ \hat{\bar{\mu}}_\alpha(t), \hat{O}_\mathrm{S}(t) \right] \left[ e^{-i (\omega + \omega_\beta) (t-t')}  \hat{\sigma}^{(+)}_\beta(t) + e^{-i (\omega - \omega_\beta) (t-t')}  \hat{\sigma}^{(-)}_\beta(t) \right] \right\} + \hat{C}_\mathrm{free}(t) 
\label{Eq:Commutator_MPh0_3}\\
\nonumber
    & \quad = - \sum_{\alpha, \beta} \int_0^{\infty} \dd{\omega} J_{0, \alpha \beta}(\omega) \left\{  \left[ \left( \pi \delta(\omega - \omega_\beta) + i \mathcal{P} \frac{1}{\omega - \omega_\beta} \right)  \hat{\sigma}^{(+)}_\beta(t) + \left( \pi \delta(\omega + \omega_\beta) + i \mathcal{P} \frac{1}{\omega + \omega_\beta} \right)  \hat{\sigma}^{(-)}_\beta(t) \right] \left[ \hat{\bar{\mu}}_\alpha(t), \hat{O}_\mathrm{S}(t) \right] \right. \\
    & \quad \quad \left. -  \left[ \hat{\bar{\mu}}_\alpha(t), \hat{O}_\mathrm{S}(t) \right] \left[ \left( \pi \delta(\omega + \omega_\beta) - i \mathcal{P} \frac{1}{\omega + \omega_\beta} \right)  \hat{\sigma}^{(+)}_\beta(t) + \left( \pi \delta(\omega - \omega_\beta) - i \mathcal{P} \frac{1}{\omega - \omega_\beta} \right)  \hat{\sigma}^{(-)}_\beta(t) \right] \right\} + \hat{C}_\mathrm{free}(t) 
\label{Eq:Commutator_MPh0_4} \\
\nonumber
    & \quad \approx - \sum_{\alpha, \beta} \left\{ \left[ \pi J_{0, \alpha \beta}(\omega_\beta) + i \delta^{0-}_{\alpha\beta}(\omega_\beta) \right] \left[ \hat{\sigma}^{(+)}_\beta(t) \hat{\sigma}^{(-)}_\alpha(t) \hat{O}_\mathrm{S}(t) - \hat{\sigma}^{(+)}_\beta(t) \hat{O}_\mathrm{S}(t) \hat{\sigma}^{(-)}_\alpha(t) \right] \right. \\
\nonumber
    & \quad \quad \quad + i \delta^{0+}_{\alpha\beta}(\omega_\beta) \left[ \hat{\sigma}^{(-)}_\beta(t) \hat{\sigma}^{(+)}_\alpha(t) \hat{O}_\mathrm{S}(t) - \hat{\sigma}^{(-)}_\beta(t) \hat{O}_\mathrm{S}(t) \hat{\sigma}^{(+)}_\alpha(t) \right] + i \delta^{0+}_{\alpha\beta}(\omega_\beta) \left[ \hat{\sigma}^{(-)}_\alpha(t) \hat{O}_\mathrm{S}(t) \hat{\sigma}^{(+)}_\beta(t) - \hat{O}_\mathrm{S}(t) \hat{\sigma}^{(-)}_\alpha(t) \hat{\sigma}^{(+)}_\beta(t) \right] \\
    & \quad \quad \quad \left. - \left[ \pi J_{0, \alpha \beta}(\omega_\beta) - i \delta^{0-}_{\alpha\beta}(\omega_\beta) \right] \left[ \hat{\sigma}^{(+)}_\alpha(t) \hat{O}_\mathrm{S}(t) \hat{\sigma}^{(-)}_\beta(t) - \hat{O}_\mathrm{S}(t) \hat{\sigma}^{(+)}_\alpha(t) \hat{\sigma}^{(-)}_\beta(t) \right] \right\} + \hat{C}_\mathrm{free}(t).
\label{Eq:Commutator_MPh0_5}
\end{align}
From Eq.~(\ref{Eq:Commutator_MPh0_3}) to (\ref{Eq:Commutator_MPh0_4}), we have utilized the identity $\int_{-\infty}^t \dd{t'} e^{\pm i(\omega + \omega_\beta) (t-t')} = \int_0^\infty \dd{\tau} e^{\pm i(\omega + \omega_\beta)\tau} = \pi \delta(\omega + \omega_\beta) \pm i \mathcal{P} \frac{1}{\omega + \omega_\beta}$; from Eq.~(\ref{Eq:Commutator_MPh0_4}) to (\ref{Eq:Commutator_MPh0_5}), we have dropped the terms containing double excitation, i.e., $\hat{\sigma}^{(+)}_{\alpha (\beta)}(t) \hat{\sigma}^{(+)}_{\alpha (\beta)}(t)$, and double de-excitation, i.e., $\hat{\sigma}^{(-)}_{\alpha (\beta)}(t) \hat{\sigma}^{(-)}_{\alpha (\beta)}(t)$. Finally, using the identities,
\begin{align}
\nonumber
    \begin{cases}
        \hat{\sigma}^{(-)}_{\alpha} \hat{\sigma}^{(+)}_{\alpha} + \hat{\sigma}^{(+)}_{\alpha} \hat{\sigma}^{(-)}_{\alpha} = \hat{\mathrm{I}}_\alpha,  \\
        \hat{\sigma}^{(-)}_{\alpha} \hat{\sigma}^{(+)}_{\beta} = \hat{\sigma}^{(+)}_{\beta} \hat{\sigma}^{(-)}_{\alpha}, ~~\alpha \neq \beta, 
    \end{cases}
\end{align}
Eq.~(\ref{Eq:Commutator_MPh0_5}) can be expressed as
\begin{align}
\nonumber
    & \frac{i}{\hbar} \left[ \hat{\mathcal{H}}^0_\mathrm{ES}(t), \hat{O}_\mathrm{S}(t) \right] + \sum_{\alpha, \beta} \Gamma^0_{\alpha \beta} \left[ \hat{\sigma}^{(+)}_\alpha(t) \hat{O}_\mathrm{S}(t) \hat{\sigma}^{(-)}_\beta(t) - \frac{1}{2} \left\{ \hat{\sigma}^{(+)}_\alpha(t) \hat{\sigma}^{(-)}_\beta(t), \hat{O}_\mathrm{S}(t) \right\} \right] \\
    & \quad + \sum_{\alpha, \beta} \gamma^0_{\alpha \beta} \left[ \hat{\sigma}^{(-)}_\alpha(t) \hat{O}_\mathrm{S}(t) \hat{\sigma}^{(+)}_\beta(t) - \frac{1}{2} \left\{ \hat{\sigma}^{(-)}_\alpha(t) \hat{\sigma}^{(+)}_\beta(t), \hat{O}_\mathrm{S}(t) \right\} \right] + \hat{C}_\mathrm{free}(t).
\label{Eq:Commutator_MPh0_6}
\end{align}

We then move forward to the second term on the right-hand side of Eq.~(\ref{Eq:Commutator_SB}), which can be expanded as
\begin{align}
    \frac{i}{\hbar} \left[ \hat{\mathcal{H}}_\mathrm{Ph1-Ph2}(t), \hat{O}_\mathrm{S}(t) \right] = i \sum_j \int_{-\infty}^\infty \dd{\omega} \sqrt{\frac{\kappa_{\mathrm{ph}, j}}{2\pi}} \left\{ \hat{b}_j^\dagger(\omega,t) \left[ \hat{a}_j(t), \hat{O}_\mathrm{S}(t) \right] + \left[ \hat{a}_j^\dagger(t), \hat{O}_\mathrm{S}(t) \right] \hat{b}_j(\omega,t) \right\}.
\label{Eq:Commutator_Ph1Ph2}
\end{align}
$\hat{b}_j(\omega, t)$ and $\hat{b}^\dagger_j(\omega, t)$ can again be solved from the Heisenberg equation, resulting in
\begin{subequations}
\begin{gather}
    \hat{b}_j(\omega, t) = \hat{b}_{j, \mathrm{free}}(\omega, t) - i \sqrt{\frac{\kappa_{\mathrm{ph}, j}}{2\pi}} \int_0^t \dd{t'} e^{-i \omega (t-t')} \hat{a}_j(t'), 
\label{Eq:b_dyanmics} \\
    \hat{b}^\dagger_j(\omega, t) = \hat{b}^\dagger_{j, \mathrm{free}}(\omega, t) + i \sqrt{\frac{\kappa_{\mathrm{ph}, j}}{2\pi}} \int_0^t \dd{t'} e^{i \omega (t-t')} \hat{a}^\dagger_j(t'),
\label{Eq:b_dagger_dyanmics}
\end{gather}    
\end{subequations}
where $\hat{b}_{j, \mathrm{free}}(\omega, t) = e^{-i \omega t} \hat{b}_j(\omega, 0)$ and $\hat{b}^\dagger_{j, \mathrm{free}}(\omega, t) = e^{i \omega t} \hat{b}^\dagger_j(\omega, 0)$ are the free-evolution of $\hat{b}_j(\omega, t)$ and $\hat{b}^\dagger_j(\omega, t)$. Plugging Eqs.~(\ref{Eq:b_dyanmics}) and (\ref{Eq:b_dagger_dyanmics}) into Eq.(\ref{Eq:Commutator_Ph1Ph2}), we obtain
\begin{align}
    & - \sum_j \frac{\kappa_{\mathrm{ph}, j}}{2\pi} \int_0^t \dd{t'} \int_{-\infty}^{\infty} \dd{\omega} \left\{ e^{i \omega (t-t')} \hat{a}^\dagger_j(t') \left[ \hat{a}_j(t), \hat{O}_\mathrm{S}(t) \right] - \left[ \hat{a}^\dagger_j(t), \hat{O}_\mathrm{S}(t) \right] e^{-i \omega (t-t')} \hat{a}_j(t') \right\} + \hat{B}_\mathrm{free}(t),
\label{Eq:Commutator_Ph1Ph2_2}
\end{align}
where
\begin{align}
    \hat{B}_\mathrm{free}(t) = i \sum_j \sqrt{\frac{\kappa_{\mathrm{ph}, j}}{2\pi}} \int_{-\infty}^{\infty} \dd{\omega} \left\{ \hat{b}^\dagger_{j, \mathrm{free}}(\omega, t) \left[ \hat{a}_j(t), \hat{O}_\mathrm{S}(t) \right] + \left[ \hat{a}^\dagger_j(t), \hat{O}_\mathrm{S}(t) \right] \hat{b}_{j, \mathrm{free}}(\omega, t) \right\}.
\end{align}
Using the identity $\int_{-\infty}^{\infty} \dd{\omega} e^{\pm i \omega (t-t')} = 2 \pi \delta(t-t')$, Eq.~(\ref{Eq:Commutator_Ph1Ph2_2}) can be transformed to 
\begin{align}
\nonumber
    & - \sum_j \frac{\kappa_{\mathrm{ph}, j}}{2} \left\{ \hat{a}^\dagger_j(t) \left[ \hat{a}_j(t), \hat{O}_\mathrm{S}(t) \right] - \left[ \hat{a}^\dagger_j(t), \hat{O}_\mathrm{S}(t) \right] \hat{a}_j(t) \right\} + \hat{B}_\mathrm{free}(t) \\
    & \quad = \sum_j \kappa_{\mathrm{ph}, j} \left[ \hat{a}^\dagger_j(t) \hat{O}_\mathrm{S}(t) \hat{a}_j(t) - \frac{1}{2} \left\{ \hat{a}^\dagger_j(t) \hat{a}_j(t), \hat{O}_\mathrm{S}(t) \right\} \right] + \hat{B}_\mathrm{free}(t).
\label{Eq:Commutator_Ph1Ph2_3}
\end{align}
Combining Eqs.~(\ref{Eq:Commutator_OS}), (\ref{Eq:Commutator_SB}), (\ref{Eq:Commutator_MPh0_6}) and (\ref{Eq:Commutator_Ph1Ph2_3}), we obtain
\begin{align}
\nonumber
    & \pdv{t} \hat{O}_\mathrm{S}(t) = \\
\nonumber    
    & \quad \frac{i}{\hbar} \left[ \hat{\mathcal{H}}_\mathrm{S}(t) + \hat{\mathcal{H}}^0_\mathrm{ES}(t), \hat{O}(t) \right] + \sum_{\alpha, \beta} \Gamma^0_{\alpha \beta} \left[ \hat{\sigma}^{(+)}_\alpha(t) \hat{O}_\mathrm{S}(t) \hat{\sigma}^{(-)}_\beta(t) - \frac{1}{2} \left\{ \hat{\sigma}^{(+)}_\alpha(t) \hat{\sigma}^{(-)}_\beta(t), \hat{O}_\mathrm{S}(t) \right\} \right] \\
\nonumber
    & \quad + \sum_{\alpha, \beta} \gamma^0_{\alpha \beta} \left[ \hat{\sigma}^{(-)}_\alpha(t) \hat{O}_\mathrm{S}(t) \hat{\sigma}^{(+)}_\beta(t) - \frac{1}{2} \left\{ \hat{\sigma}^{(-)}_\alpha(t) \hat{\sigma}^{(+)}_\beta(t), \hat{O}_\mathrm{S}(t) \right\} \right]  + \sum_j \kappa_{\mathrm{ph}, j} \left[ \hat{a}^\dagger_j(t) \hat{O}_\mathrm{S}(t) \hat{a}_j(t) - \frac{1}{2} \left\{ \hat{a}^\dagger_j(t) \hat{a}_j(t), \hat{O}_\mathrm{S}(t) \right\} \right] \\
    & \quad + \hat{B}_\mathrm{free}(t) + \hat{C}_\mathrm{free}(t).
\end{align}
We here consider that the bath (photonic) modes, i.e., $\hat{b}_j(\omega)$ and $\hat{c}_l(\omega)$, are initially in the vacuum state, thus,
\begin{align}
    \left< \hat{B}_\mathrm{free}(t) \right> = \left< \hat{C}_\mathrm{free}(t) \right> = 0,
\end{align}
and
\begin{align}
\nonumber
    & \pdv{t} \left< \hat{O}_\mathrm{S}(t) \right> = \\
\nonumber    
    & \quad \biggl< \frac{i}{\hbar} \left[ \hat{\mathcal{H}}_\mathrm{S}(t) + \hat{\mathcal{H}}^0_\mathrm{ES}(t), \hat{O}_\mathrm{S}(t) \right] + \sum_{\alpha, \beta} \Gamma^0_{\alpha \beta} \left[ \hat{\sigma}^{(+)}_\alpha(t) \hat{O}_\mathrm{S}(t) \hat{\sigma}^{(-)}_\beta(t) - \frac{1}{2} \left\{ \hat{\sigma}^{(+)}_\alpha(t) \hat{\sigma}^{(-)}_\beta(t), \hat{O}_\mathrm{S}(t) \right\} \right] \\
    & \quad + \sum_{\alpha, \beta} \gamma^0_{\alpha \beta} \left[ \hat{\sigma}^{(-)}_\alpha(t) \hat{O}_\mathrm{S}(t) \hat{\sigma}^{(+)}_\beta(t) - \frac{1}{2} \left\{ \hat{\sigma}^{(-)}_\alpha(t) \hat{\sigma}^{(+)}_\beta(t), \hat{O}_\mathrm{S}(t) \right\} \right] + \sum_j \kappa_{\mathrm{ph}, j} \left[ \hat{a}^\dagger_j(t) \hat{O}_\mathrm{S}(t) \hat{a}_j(t) - \frac{1}{2} \left\{ \hat{a}^\dagger_j(t) \hat{a}_j(t), \hat{O}_\mathrm{S}(t) \right\} \right] \biggr>.
\label{Eq:Os_expectation}
\end{align}
Note that even when the molecules are in a thermal environment, where the bath modes are not initially in the vacuum state, the dynamics of the molecules can still be equivalently described by a temperature-dependent generalized spectral density with the initial state of the environmental (bath) modes being the vacuum state \cite{Tamascelli2019}.

Finally, recalling the relationship
\begin{align}
\nonumber
    \left< \hat{O}_\mathrm{S}(t) \right> = \mathrm{Tr_S} \mathrm{Tr_B} \left[ \hat{O}_\mathrm{S}(t) \hat{\rho}_\mathrm{T}(0) \right] = \mathrm{Tr_S} \left[ \hat{O}_\mathrm{S}(t) \hat{\rho}_\mathrm{S}(0) \right] = \mathrm{Tr_S} \left[ \hat{O}_\mathrm{S} \hat{\rho}_\mathrm{S}(t) \right]
\end{align}
($\hat{\rho}_\mathrm{T}$ is the total density matrix of the system and bath, and $\hat{\rho}_\mathrm{S}$ is the reduced density matrix of the system) and using the cyclic rule of trace, the left-hand side of Eq.~(\ref{Eq:Os_expectation}) can be written as
\begin{align}
    \pdv{t} \left< \hat{O}_\mathrm{S}(t) \right> = \mathrm{Tr_S} \left\{ \left[ \pdv{t} \hat{O}_\mathrm{S}(t) \right] \hat{\rho}_\mathrm{S}(0) \right\} = \mathrm{Tr_S} \left\{ \hat{O}_\mathrm{S} \left[ \pdv{t} \hat{\rho}_\mathrm{S}(t) \right] \right\},
\label{Eq:Os_expectation_left}
\end{align}
and the right-hand side of Eq.~(\ref{Eq:Os_expectation}) can be written as
\begin{align}
\nonumber
    & \mathrm{Tr_S} \Biggl[ \hat{O}_\mathrm{S} \biggl\{ -\frac{i}{\hbar} \left[ \hat{\mathcal{H}}_\mathrm{S} + \hat{\mathcal{H}}^0_\mathrm{ES}, \hat{\rho}_\mathrm{S}(t) \right] + \sum_{\alpha, \beta} \Gamma^0_{\alpha \beta} \left[ \hat{\sigma}^{(-)}_\beta \hat{\rho}_\mathrm{S}(t) \hat{\sigma}^{(+)}_\alpha - \frac{1}{2} \left\{ \hat{\sigma}^{(+)}_\alpha \hat{\sigma}^{(-)}_\beta, \hat{\rho}_\mathrm{S}(t) \right\} \right] \\
    & \quad \quad \quad + \sum_{\alpha, \beta} \gamma^0_{\alpha \beta} \left[ \hat{\sigma}^{(+)}_\beta \hat{\rho}_\mathrm{S}(t) \hat{\sigma}^{(-)}_\alpha - \frac{1}{2} \left\{ \hat{\sigma}^{(-)}_\alpha \hat{\sigma}^{(+)}_\beta, \hat{\rho}_\mathrm{S}(t) \right\} \right] + \sum_j \kappa_{\mathrm{ph}, j} \left[ \hat{a}_j \hat{\rho}_\mathrm{S}(t) \hat{a}^\dagger_j - \frac{1}{2} \left\{ \hat{a}^\dagger_j \hat{a}_j, \hat{\rho}_\mathrm{S}(t) \right\} \right] \biggr\} \Biggr].
\label{Eq:Os_expectation_right}
\end{align}
Comparing Eqs.~(\ref{Eq:Os_expectation_left}) and (\ref{Eq:Os_expectation_right}), we obtain Eq.~(\ref{Eq:EOM_rho_CQED}) as
\begin{align}
\nonumber
    \pdv{t} \hat{\rho}_\mathrm{S}(t) & = -\frac{i}{\hbar} \left[ \hat{\mathcal{H}}_\mathrm{S} + \hat{\mathcal{H}}^0_\mathrm{ES}, \hat{\rho}_\mathrm{S}(t) \right] + \sum_{\alpha, \beta} \Gamma^0_{\alpha \beta} \left[ \hat{\sigma}^{(-)}_\beta \hat{\rho}_\mathrm{S}(t) \hat{\sigma}^{(+)}_\alpha - \frac{1}{2} \left\{ \hat{\sigma}^{(+)}_\alpha \hat{\sigma}^{(-)}_\beta, \hat{\rho}_\mathrm{S}(t) \right\} \right] \\
\nonumber
    & \quad + \sum_{\alpha, \beta} \gamma^0_{\alpha \beta} \left[ \hat{\sigma}^{(+)}_\beta \hat{\rho}_\mathrm{S}(t) \hat{\sigma}^{(-)}_\alpha - \frac{1}{2} \left\{ \hat{\sigma}^{(-)}_\alpha \hat{\sigma}^{(+)}_\beta, \hat{\rho}_\mathrm{S}(t) \right\} \right] + \sum_j \kappa_{\mathrm{ph}, j} \left[ \hat{a}_j \hat{\rho}_\mathrm{S}(t) \hat{a}^\dagger_j - \frac{1}{2} \left\{ \hat{a}^\dagger_j \hat{a}_j, \hat{\rho}_\mathrm{S}(t) \right\} \right].
\end{align}

\renewcommand\arraystretch{1.5}
\begin{table*}[h]
    \centering
    \resizebox{\columnwidth}{!}{\begin{tabular}{c||cc||cccccccccccccccc}
    \toprule
        FIG. & $h$ (nm) & $d$ (nm) & $\hbar\omega_{\mathrm{ph}, 1}$ & $\hbar\omega_{\mathrm{ph}, 2}$ & $\hbar\omega_{\mathrm{ph}, 3}$ & $\hbar\omega_{\mathrm{ph}, 4}$ & $\hbar\kappa_{\mathrm{ph}, 1}$ & $\hbar\kappa_{\mathrm{ph}, 2}$ & $\hbar\kappa_{\mathrm{ph}, 3}$ & $\hbar\kappa_{\mathrm{ph}, 4}$ & $\hbar\Omega_{11}$ & $\hbar\Omega_{12}$ & $\hbar\Omega_{13}$ & $\hbar\Omega_{14}$ & $\hbar\Omega_{21}$ & $\hbar\Omega_{22}$ & $\hbar\Omega_{23}$ & $\hbar\Omega_{24}$ \\
    \toprule
        \ref{Fig4}(a) & 7 & $\diagdown$ & 3.486 & 3.527 & $\diagdown$ & $\diagdown$ & 144.8 & 98.0 & $\diagdown$ & $\diagdown$ & 3.0 & 5.6 & $\diagdown$ & $\diagdown$ & $\diagdown$ & $\diagdown$ & $\diagdown$ & $\diagdown$ \\
    \hline
        \ref{Fig4}(b) & 1 & $\diagdown$ & 3.513 & 3.535 & $\diagdown$ & $\diagdown$ & 106.4 & 99.9 & $\diagdown$ & $\diagdown$ & 10.0 & 117.0 & $\diagdown$ & $\diagdown$ & $\diagdown$ & $\diagdown$ & $\diagdown$ & $\diagdown$ \\
    \hline
        \ref{Fig5}(a) & 7 & 1.5 & 3.439 & 3.499 & 3.527 & 3.530 & 192.7 & 103.9 & 97.2 & 97.2 & 1.8 & 3.0 & 3.9 & 3.7 & 1.8 & 2.9 & 5.0 & 2.0 \\
    \hline
        \ref{Fig5}(b) & 7 & 3 & 3.435 & 3.498 & 3.530 & 3.531 & 194.5 & 104.3 & 97.3 & 101.2 & 1.8 & 2.8 & 5.1 & -2.0 & 1.8 & 2.8 & 5.1 & 2.0 \\
    \hline
        \ref{Fig5}(c) & 7 & 10 & 3.408 & 3.483 & 3.523 & 3.528 & 215.4 & 111.8 & 99.2 & 102.1 & 1.4 & 2.0 & 3.8 & 4.5 & 1.4 & 2.0 & 3.8 & -4.5 \\
    \hline
        \ref{Fig5}(d) & 1 & 1.5 & 3.551 & 3.534 & 3.534 & 3.536 & 133.6 & 100.2 & 100.1 & 98.9 & -9.5 & 93.1 & -42.3 & -57.0 & 9.9 & 65.3 & 79.4 & -56.0 \\
    \hline
        \ref{Fig5}(e) & 1 & 3 & 3.516 & 3.533 & 3.535 & 3.537 & 127.7 & 98.5 & 100.0 & 100.7 & -11.3 & 62.0 & -80.2 & 58.2 & -10.4 & 47.4 & 99.1 & 40.3 \\
    \hline
        \ref{Fig5}(f) & 1 & 10 & 3.495 & 3.530 & 3.535 & 3.535 & 150.6 & 99.9 & 100.0 & 99.9 & 5.0 & 37.0 & -30.1 & 107.2 & 4.4 & 6.3 & 113.9 & 27.8 \\
    \hline
    \hline
    \end{tabular}}
    \caption{Parameters in dissipative CQED-DDI. $\hbar\omega_{\mathrm{ph}, j}$ are in eV; $\hbar\kappa_{\mathrm{ph}, j}$ and $\hbar\Omega_{\alpha j}$ are in meV.}
    \label{Tab:Parameters_single}
\end{table*}   
\renewcommand\arraystretch{1}

\renewcommand\arraystretch{1.5}
\begin{table*}[ht]
    \centering
    \resizebox{\columnwidth}{!}{\begin{tabular}{c||cc||cccccccccccccccc}
    \toprule
        FIG. & $h$ (nm) & $d$ (nm) & $\hbar\omega'_{\mathrm{ph}, 1}$ & $\hbar\omega'_{\mathrm{ph}, 2}$ & $\hbar\omega'_{\mathrm{ph}, 3}$ & $\hbar\omega'_{\mathrm{ph}, 4}$ & $\hbar\kappa'_{\mathrm{ph}, 1}$ & $\hbar\kappa'_{\mathrm{ph}, 2}$ & $\hbar\kappa'_{\mathrm{ph}, 3}$ & $\hbar\kappa'_{\mathrm{ph}, 4}$ & $\hbar\Omega'_{11}$ & $\hbar\Omega'_{12}$ & $\hbar\Omega'_{13}$ & $\hbar\Omega'_{14}$ & $\hbar\Omega'_{21}$ & $\hbar\Omega'_{22}$ & $\hbar\Omega'_{23}$ & $\hbar\Omega'_{24}$ \\
    \toprule
        \ref{Fig4}(a) & 7 & $\diagdown$ & 3.487 & 3.527 & $\diagdown$ & $\diagdown$ & 146.3 & 97.9 & $\diagdown$ & $\diagdown$ & 3.0 & 5.6 & $\diagdown$ & $\diagdown$ & $\diagdown$ & $\diagdown$ & $\diagdown$ & $\diagdown$ \\
    \hline
        \ref{Fig4}(b) & 1 & $\diagdown$ & 3.513 & 3.535 & $\diagdown$ & $\diagdown$ & 106.5 & 99.9 & $\diagdown$ & $\diagdown$ & 10.0 & 117.0 & $\diagdown$ & $\diagdown$ & $\diagdown$ & $\diagdown$ & $\diagdown$ & $\diagdown$ \\
    \hline
        \ref{Fig5}(a) & 7 & 1.5 & 3.448 & 3.502 & 3.530 & 3.530 & 201.2 & 104.3 & 96.5 & 96.5 & 1.9 & 3.1 & 4.3 & 3.0 & 1.9 & 3.1 & 2.8 & 4.4 \\
    \hline
        \ref{Fig5}(b) & 7 & 3 & 3.446 & 3.500 & 3.530 & 3.531 & 203.6 & 104.8 & 96.4 & 100.9 & 1.9 & 3.0 & 4.9 & 2.0 & 1.9 & 3.0 & 4.9 & -2.0 \\
    \hline
        \ref{Fig5}(c) & 7 & 10 & 3.425 & 3.487 & 3.524 & 3.528 & 236.0 & 112.9 & 97.7 & 102.0 & 1.6 & 2.2 & 3.6 & 4.5 & 1.6 & 2.2 & 3.6 & -4.5 \\
    \hline
        \ref{Fig5}(d) & 1 & 1.5 & 3.506 & 3.531 & 3.535 & 3.535 & 138.2 & 99.5 & 100.0 & 99.9 & -6.7 & 38.8 & -84.9 & 70.9 & -6.8 & 38.9 & 84.5 & 71.4 \\
    \hline
        \ref{Fig5}(e) & 1 & 3 & 3.513 & 3.533 & 3.535 & 3.536 & 122.5 & 99.1 & 100.0 & 100.2 & 9.5 & 68.5 & -47.2 & 82.4 & 8.2 & 23.9 & 113.9 & 13.6 \\
    \hline
        \ref{Fig5}(f) & 1 & 10 & 3.510 & 3.535 & 3.535 & 3.543 & 142.0 & 99.8 & 100.0 & 101.8 & 7.0 & 81.5 & -83.3 & 12.4 & 7.0 & 80.9 & 83.9 & 12.3 \\
    \hline
    \hline
    \end{tabular}}
    \caption{Parameters in dissipative CQED. $\hbar\omega'_{\mathrm{ph}, j}$ are in eV; $\hbar\kappa'_{\mathrm{ph}, j}$ and $\hbar\Omega'_{\alpha j}$ are in meV.}
    \label{Tab:Parameters_CQED}
\end{table*}   
\renewcommand\arraystretch{1}

\section{Parameters in dissipative CQED-DDI and dissipative CQED}
\label{Appendix:Parameters}

The parameters $\omega_{\mathrm{ph}, j}$, $\kappa_{\mathrm{ph}, j}$ and $\Omega_{\alpha j}$ used in dissipative CQED-DDI are shown in Tab.~\ref{Tab:Parameters_single}; the parameters $\omega'_{\mathrm{ph}, j}$, $\kappa'_{\mathrm{ph}, j}$ and $\Omega'_{\alpha j}$ used in dissipative CQED are shown in Tab.~\ref{Tab:Parameters_CQED}. Note that we use two modes, i.e., $N_\mathrm{ph},~N'_\mathrm{ph} = 2$, for the single-molecule cases in FIGs.~\ref{Fig4}(a) and \ref{Fig4}(b) and use four modes, i.e., $N_\mathrm{ph},~N'_\mathrm{ph} = 4$, for the two-molecule cases in FIGs.~\ref{Fig5}(a) to \ref{Fig5}(f).

\end{widetext}

\bigbreak


%

\end{document}